\documentclass{acm_proc_article-sp}

\newcommand{\paperTitle}{Intelligent Transaction Scheduling via Conflict Prediction in OLTP DBMS}
\newcommand{\paperKeywords}{}
\newcommand{\paperAuthors}{}

\usepackage{marginnote}
\usepackage{subfigure}
\usepackage{amsmath}
\usepackage{amssymb}
\usepackage{times}
\usepackage{boxedminipage}
\usepackage{xspace}
\usepackage{array}
\usepackage{epsfig}
\usepackage{calc}
\usepackage{multirow}
\usepackage{rotating}
\usepackage{enumitem}
\usepackage[usenames,dvipsnames]{color}
\usepackage{tabularx}
\usepackage{balance}  % for  \balance command ON LAST PAGE  (only there!)
\usepackage{tikz}
\usepackage{wasysym }
\usepackage{xcolor}
\usepackage[font={small}]{caption}
\usepackage{alltt}
\usepackage{setspace}
\newcommand{\subparagraph}{} % for titlesec + sig-alternate
\usepackage{etoolbox}

\usepackage{times}
\usepackage[final]{microtype}
\usepackage[scaled]{inconsolata}
\usepackage[T1]{fontenc}
\usepackage{graphicx}
\usepackage{soul}
\usepackage{pifont}

\usepackage[
            bookmarks=true, 
            bookmarksopen=true, 
            pdfhighlight=/I,
            pdfpagemode=UseOutlines, 
            linkcolor=blue, 
            pdfborder={ 0 0 0 },
            pageanchor=false]{hyperref}
\hypersetup{
 pdfauthor = {\paperAuthors},
 pdftitle = {\paperTitle},
 pdfkeywords = {\paperKeywords},
 pdfborder={ 0 0 0 }
}

\usepackage[hyphenbreaks]{breakurl}
\usepackage[capitalize,noabbrev,nameinlink]{cleveref}

\newcommand{\peloton}{Peloton\xspace}

\definecolor{comment-red}{rgb}{1,0,0}

\newcommand{\squishitemize}{
 \begin{list}{$\bullet$}
  { \setlength{\itemsep}{0pt}
     \setlength{\parsep}{3pt}
     \setlength{\topsep}{3pt}
     \setlength{\leftmargin}{1.95em}
     \setlength{\labelwidth}{1.5em}
     \setlength{\labelsep}{0.5em} } }

\newcounter{Lcount}
\newcommand{\squishlist}{
    \begin{list}{\arabic{Lcount}. }
   { \usecounter{Lcount}
        \setlength{\itemsep}{0pt}
        \setlength{\parsep}{3pt}
        \setlength{\topsep}{3pt}
        \setlength{\partopsep}{0pt}
        \setlength{\leftmargin}{2em}
        \setlength{\labelwidth}{1.5em}
        \setlength{\labelsep}{0.5em} } }

\newcommand{\squishend}{\end{list}}

\begin{document}

\title{\paperTitle}
%\subtitle{[Extended Abstract]}

%\numberofauthors{3} 
%\author{
%\alignauthor
%Tieying Zhang\\
%      \affaddr{Bytedance}\\
%      \email{tieying.zhang@bytedance.com}
%\alignauthor
%Anthony Tomasic\\
%      \affaddr{Carnegie Mellon University}\\
%      \email{tomasic@cs.cmu.edu}
%\alignauthor
%Andrew Pavlo\\
%      \affaddr{Carnegie Mellon University}\\
%      \email{pavlo@cs.cmu.edu}
%}

\author{
  Tieying Zhang\\
  \texttt{Bytedance}\\
  \texttt{tieying.zhang@bytedance.com}
  \and
  Anthony Tomasic\\
  \texttt{Carnegie Mellon University}\\
  \texttt{tomasic@cs.cmu.edu}
  \and
  Andrew Pavlo\\
  \texttt{Carnegie Mellon University}\\
  \texttt{pavlo@cs.cmu.edu}
}

\maketitle

%% ==================================================================
%% abstract
%% ==================================================================
\begin{abstract}
Current architectures for main-memory online transaction processing (OLTP) database management systems (DBMS) typically use random scheduling to assign transactions to threads. This approach achieves uniform load across threads but it ignores the likelihood of conflicts between transactions.  If the DBMS could estimate the potential for transaction conflict and then intelligently schedule transactions to avoid conflicts, then the system could improve its performance. Such estimation of transaction conflict, however, is non-trivial for several reasons. First, conflicts occur under complex conditions that are far removed in time from the scheduling decision.  Second, transactions must be represented in a compact and efficient manner to allow for fast conflict detection. Third, given some evidence of potential conflict, the DBMS must schedule transactions in such a way that minimizes this conflict. In this paper, we systematically explore the design decisions for solving these problems. We then empirically measure the performance impact of different representations on standard OLTP benchmarks. Our results show that intelligent scheduling using a history increases throughput by $\sim$40\% on 20-core machine.

\end{abstract}

%A category including the fourth, optional field follows...
%\category{D.2.8}{Software Engineering}{Metrics}[complexity measures, performance measures]


%\keywords{transaction scheduling, statistical modeling} % NOT required for Proceedings

%% ==================================================================
%% Introduction
%% ==================================================================
\section{Introduction}

%Database administrators (DBAs) use various techniques to balance load and reduce transaction conflicts. For example, DBAs schedule maintenance operations during off hours, or partition the data~\cite{scalzo2003oracle}. Typically a DBA uses informal knowledge of the workload, data, and schema to accomplish the goal of packing more transactions per second into a database system~\cite{thiele2009partition,ahmad2011predicting}. Unfortunately, these decisions by DBAs may be incorrect, inefficient, arrive too late or too infrequently. {\em Autonomous databases}~\cite{Pavlo:2017} is a new research area that aims to replace DBA decisions with autonomous decision making systems. The long-term goal of this research area is to construct database systems that are easier to use, cheaper, and more efficient. In this paper, we study one DBA decision: scheduling transactions to avoid conflicts.

A significant source of performance degradation in OLTP DBMS is transaction conflict on shared data. Research on this problem has led to many variations of optimistic concurrency control (OCC) and pessimistic concurrency control (e.g., two-phase locking -- 2PL) protocols.

One OLTP architectural technique to deal with this problem is to partition the data to reduce conflicts~\cite{weldon2013data, Stonebraker:2007:EAE:1325851.1325981, Thomson:2012:SIGMOD}. Adopting partitioning as a core assumption, along with several additional design decisions, achieves high performance.
However, the partitioning assumption comes at a cost. Some database schemas are not easily partitionable~\cite{Pavlo:2012:SAD:2213836.2213844,Curino:2010:SWA:1920841.1920853}. Changing the partition as the schema and data evolve is cumbersome, especially with modern large-scale, continuously operating  applications. Small mismatches between the partitioning and the data lead to poor performance. 

Another OLTP architectural technique instead focuses on careful implementation of ACID transaction properties to achieve high performance~\cite{Tu:2013:STM:2517349.2522713,kim2016ermia,yu2016tictoc}. Systems based on this architecture approach offer good performance over all database schemas and do not suffer from the cumbersome constraints of partitioning. However, performance does not match partitioned databases under all workloads.

These two architecture techniques are at two ends of a spectrum. We hypothesize a third approach that may offer the advantages of both: statistical scheduling of transactions. The intuition here is to start with a unpartitioned OLTP system and then predict conflicts and schedule transactions such that (statistically) the system achieves an abort rate that (nearly) matches a perfectly partitioned system.
To understand this better, we present a empirical exploration of reducing transaction aborts through a scheduling model based on a statistical analysis of the workload.

%The hypothesis of this paper is that this DBA decision, in some cases, can be replaced by an autonomous system that uses statistical reasoning.

\begin{figure}[t!]
    \centering
    \includegraphics[width=8.9cm]{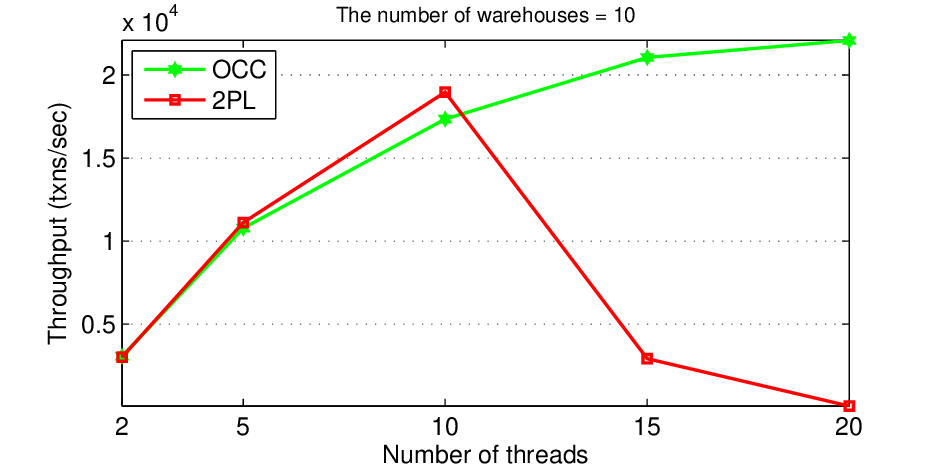}
    \caption{
        \textbf{Throughput of Randomized Scheduling} -- On a main-memory dbms the experiment measures the throughput of optimistic concurrency control (OCC) and two-phase locking (2PL) in an unthrottled system running TPC-C with a fixed number of warehouses and threads (one thread reserved for workload generation). Transactions are scheduled randomly into different threads.
%        \todo{This figure is unnecessarily large. I would combine it with Figure 2 using subfloats. Two graphs side-by-side.}
    }
    \label{fig:throughput_occ_2pl}
\end{figure}

% Generally, OCC performs well when transaction conflict rates are low and 2PL performs well when conflict rates are high.  
Consider the performance of the \peloton~\cite{Pavlo:2017-real} main-memory database system running TPC-C~\cite{TPCC} with 10 warehouses in an unthrottled experiment.\footnote{A throttled experiment regulates the arrival rate of transactions. An unthrottled experiment maximizes the throughput by continuously injecting transactions. The system is run to accept transactions as fast as possible, without a limit to the arrival rate.} 
% In an unthrottled experiment
The results in \cref{fig:throughput_occ_2pl} show that as the number of threads increases to equal the number of warehouses, throughput increases to almost 20,000 tps.
At this point, 2PL reaches a point of extreme lock contention while OCC continues to improve somewhat until the system reaches maximum throughput. 

%This is throughput of occ and 2pl \cref{fig:throughput_occ_2pl}.

% \begin{figure}
% \centering
% \includegraphics[width=8.9cm]{fig/response_time_occ_2pl.eps}
% \caption{The performance of optimistic concurrency control (OCC) and two-phase locking (2PL) in a throttled system running TPC-C. Each point on a line represents the addition of 2000 transactions/second to the throttled system, starting at 2000 and ending at 18,000.}
% \label{fig:closed-occ-2pl}
% \end{figure}

\begin{figure}
    \centering
    \includegraphics[width=8.9cm]{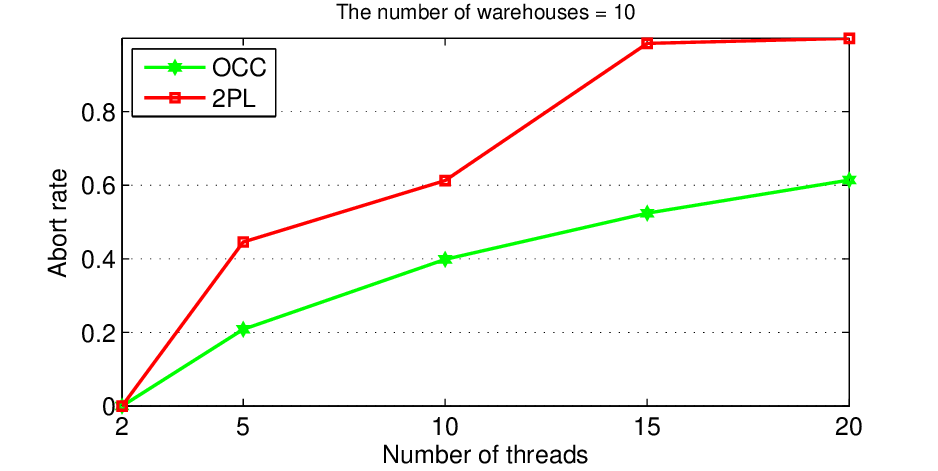}
    % trim left bottom right top
    \caption{\textbf{Abort Rate of Randomized Scheduling} -- 
        The abort rate running unthrottled on TPC-C with a fixed number of warehouses. As more threads are added to the system, the abort rate increases, because transaction are added randomly to the thread queues, without regard to potential conflicts.
    }
    \label{fig:abort-rate-occ-2pl}
\end{figure}

Transaction aborts are an essential cause of this lock contention.
As the number of active transactions in the DBMS increases, the likelihood of conflict between transactions increases (\cref{fig:abort-rate-occ-2pl}) because transactions are assigned to threads without regard to conflicts.

The key insight of this paper is simple: two transactions that are  likely to conflict should not be run concurrently. 
The crux of this idea is similar to the groundbreaking work of~\cite{Bailis:2014:VLDB,Stonebraker:2007:EAE:1325851.1325981,Thomson:2012:SIGMOD}.
However, these works deterministically prove conflicts in the current set of transactions. Our complimentary approach is to (i) model the probability of aborts in the current set of transactions to predict the transactions that are likely to abort based on the state of the system, and (ii) simply schedule in a first-in-first-out (FIFO) queue transactions that are likely to abort. Transactions that are scheduled in this way are unlikely to conflict with each other.  This approach is robust - as long as it provides some useful information, performance will improve.

The contributions of this paper are the following:
\squishlist
    \item 
    A novel example of the use of modeling to reduce transaction aborts,

    \item
    An empirical exploration of the performance of multiple designs that construct a variety of models,

    \item 
    A demonstration of the {\em robustness} of these designs, under high conflict conditions, across a range of OLTP benchmarks,

    \item
    A description of a simple-to-implement design that readily adapts to any database system and which empirically lowers transaction abort rate and improves throughput, and

    \item
    A more complex-to-implement design that empirically lowers the transaction abort rate and improves throughput, approaching the performance of partitioned databases.
\squishend

In the next sections of the paper, we describe the design challenges of our work and the architecture of the environment and of our system. We then describe the methodology in detail.  In particular we describe the basis of multiple policies for queuing transactions. The experiment framework is described in \cref{sec:framework}. This section is followed by experimental results that explores the policy space, followed by a discussion. We conclude with a discussion of related work a summary of the work in this paper.

\section{Challenges}
The main challenge for our proposal consists of modeling the complex behavior of a database engine that results in transaction aborts.
% model design (and machine learning model design by extension) consists of three parts: the data, the feature space, and the objective function. Logging every transaction abort and commit produces a wealth of data.   However, more complex feature spaces lead to sparser data since any combination of features is less likely to occur in the data.  More complex objective functions model more complex phenomena, but at a higher computational cost. 

A second challenge is to effectively {\em aggregate incoming transactions} into groups that are likely to conflict with each other.
Given these groups, the DBMS heuristically schedules transactions of each group linearly into a FIFO queue, thus insuring that the members of the group do not run concurrently and (with high probability) do not generate an abort.

The third challenge is the {\em costs} of operations in this approach: the cost of maintaining the model, the cost of grouping, and the cost of scheduling transactions. In this paper, our target processing environment is main memory databases with OLTP workloads.
%Consequentially, any additional logging, feature extraction or computation of objective functions has an impact on performance because computational resources are limited in this environment. 
Every CPU cycle spent on  our system is a cycle not spent on transaction processing.
Therefore we limit our design space to operations that are at most linear with respect to the length of a transaction statement.

%% ==================================================================
%% Design
%% ==================================================================
\section{Design} 

Our initial thought was to use the general principles of database query processing to model the system and estimate the cost of a transaction abort. Leveraging existing histogram statistics, we could estimate the read/write sets of a transaction and compute the probability of two transactions conflicting. 
However, the computations are expensive (proportional to the size of the data) and inaccurate because histograms become stale over time.  
We rejected this approach of directly modeling what the system would do and instead we used transaction execution as a source of evidence for transaction aborts. Every time a transaction commits or aborts, we log the circumstances of the abort and use this information to build our model.

Consider the SQL update statement from the TPC-C benchmark (\cref{fig:sql}). The SET and WHERE clauses of the statement provide some evidence to the circumstances of a commit or abort, in the following sense: any other transaction that references the same data is a candidate source of transaction conflict. That is, any other transaction that contains (e.g., \verb|s_i_id=2|) may conflict with the transaction of \cref{fig:sql}. In this paper,  a {\em reference} is a sequence of attribute, operator, and value (or a boolean combination) that appears in SELECT, UPDATE, INSERT, and DELETE statements.

Thus, the circumstances of a transaction is defined as the statements and input parameters of the transaction. In particular, we {\em represent a transaction by the set of its references}. Although this representation of the circumstances is not as precise as the read-write set of a transaction, the representation has the advantage of being a compact and cheap to compute.

Finally, with respect to grouping, we realized that a group represents transactions that historically have conflicted with each other, and that a key simplification would be to {\em assign a group to each thread}. 
With these insights, the basic design framework becomes clear. When a transaction arrives, first references are extracted. The system computes the best group for a transaction from these references, and the group is assigned to a queue for a thread.

Note our heuristic simplification of the problem of scheduling the current set of transaction given the probabilities of conflict: two transactions that are likely to abort are placed into the same group, and thus into the same queue of the same thread.  Two transactions in the same queue of a thread rarely run concurrently and thus rarely conflict. 
However, our experiments reveal problematic edge cases. 

\begin{figure}
    \begin{center}
    %\begin{boxed}
    \verb|UPDATE stock SET s_quantity=$1| \\
    \verb|WHERE s_i_id=$2 AND s_w_id=$3|
    %\end{boxed}
    \end{center}
    \caption{\textbf{A SQL Update Statement from the TPC-C benchmark} -- The variables \$1, \$2, and \$3 are instantiated at execution time.
    }
    \label{fig:sql}
\end{figure}

\section{System Architecture}
\begin{figure}[t!]
    \centering
    \includegraphics[width=6cm,clip,trim=1cm 0.5cm 1cm 0.5cm]{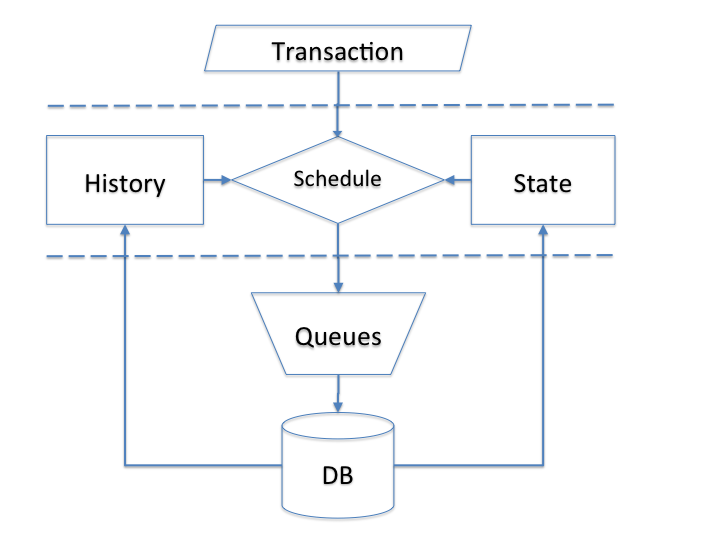}
    % trim left bottom right top
    \caption{
        \textbf{Functional System Architecture} -- 
        The Transaction component represents an incoming transaction. The History component represents the historical model of aborted transactions. The State component represents the mapping of transaction references to queues. The Schedule component chooses the queue to place the incoming transaction. The Queues component represents the set of run queues. The DB component is a main-memory database system. The dotted lines separate conventional components from the new system.
%        \\ \todo{$\uparrow$You should try to format all of the captions like this. Use a bold phrase in the beginning, followed by a '--', then the text of the caption.}
    }
\label{fig:arch}
\end{figure}

The overall environment (\cref{fig:arch}) consists of three layers -- the incoming stream of transaction requests, our system, and a main-memory database system. 
The API between the system and the database is simple, consisting of three functions: (i) the ability to capture incoming transactions, (ii) the ability to queue a transaction in a specific run queue of the database, and (iii) the ability to log the SQL text of two events that occur during transaction processing: a transaction abort, and a transaction commit. This environment and API implies our techniques can be 'bolted on' to any DBMS with little effort.

\subsection{Overview}
Our system consists of three components: (i) a \textit{History} component that extracts and aggregates information each time the system detects an abort or commit, (ii) a \textit{State} component that maps transaction references to queues, and (iii) a \textit{Schedule} component that incrementally performs the grouping operation. 
\vspace*{-0.01in} \\

% \subsection{History component}
\textbf{History Component:}
The History component is an aggregated historical summary of the references extracted from the database when a transaction aborts or commits. 
The History collects aggregated statistics depending on abort and commit events in the database. The History is implemented as a hash table. The key is reference and the value is a pair of integer counters that count the number of transactions that have aborted or committed with the associated reference. 
A counter is incremented every time a transactions aborts or commits. A counter is never decremented. The History is operational during randomized scheduling to capture the true distribution of transaction commits and aborts. During intelligent scheduling, the History component serves as a data repository.

% \subsection{State component}
\textbf{State Component:}
The State component records for each reference the number of transactions in each queue that contain the \verb|reference|. 
The State is implemented as an key-value hash table. 
The key is (\verb|reference|) and the value is an array of integers indexed by queue number. 
An array value records the number of times the associated \verb|reference| appears in the transaction queue.
The values are incremented when a transaction is queued. For each reference, we track the arrival rate per second, the total transactions in the system, and the queue with the maximum count (\cref{tab:example}).
%\vspace*{-0.01in} \\

\textbf{Schedule Component:}
Given the History and the State and an incoming transaction, it computes a score for each queue and then schedules the incoming transaction into the best queue. 
%\vspace*{-0.01in} \\

The three other components, Transaction, Queues, and DB, are standard components present in every database system. The Transaction component in the figure represents an incoming transaction. The Queues component is the set of run queues of the Database (`DB') component.

\subsection{Stealing}

In our implementation, an idle thread {\em steals} a transaction from the queue of another thread (in random fashion)~\cite{Leis:2014:MPN:2588555.2610507}.
In other words, in low utilization environments, a transaction may not be executed by the thread in which it was originally queued.
This rule prevents the thread from sitting idle. 
Of course, a stolen transaction will immediately start running concurrently with the transaction running from the non-idle thread. 
Since these transactions may conflict, this behavior is not ideal from our perspective. 
%We believe, however, that almost every DBMS implements this behavior, so we have copied this behavior to insure that our experimental results generalize to other systems.
%\todo{$\leftarrow$Do we know what the major commercial DBMSs do? - no we do not, good question.}

Note that once a transaction is assigned to a specific queue, it eventually must be executed by the thread associated with this queue, unless it is stolen. 
And once the transaction is stolen by a thread, the stealing thread always executes the transaction to completion, because aborts are immediately retried without the possibility for another thread to steal a transaction retrying after an abort. 
This behavior is not ideal from our perspective because an abort provides strong evidence that the transaction will likely abort upon re-execution under current conditions.
However, many database systems implement a variation of this abort-handling methodology. 
Our implementation thus tests our policies under the most difficult circumstances.

\subsection{Bookkeeping}
The History and State components use libcuckoo, a thread-safe and lock-free data structures~\cite{li2014algorithmic}, but the components do not globally synchronize with the Schedule component in a transactional sense.
% \todo{$\leftarrow$What does this mean? If they're thread-safe, then they can sychronize. - actually no, they can't - to be synchronized we would need transactions.}
Thus the Schedule component may make a scheduling decision based on (slightly) stale information.
This trade-off for performance over accuracy is well worth it, since the cost of synchronization is high~\cite{Tu:2013:STM:2517349.2522713}. 
The Schedule component is robust to minor inaccuracies in bookkeeping because both History and State components store an aggregated history of behavior that only slowly changes over time.
Our implementation ensures that access to these structures is not a contention point in the system.

%% ==================================================================
%% Scheduling
%% ==================================================================
\section{Policies}
\label{sec:method}

The overall process for scheduling a new arrived transaction into a queue consists of several steps: (i) extract the references from the transaction according to a {\em policy}, (ii) look up the references in the History to gather historical evidence on the conflict rates of references, (iii) produce a conflict likelihood score by combining this evidence with the historical state mapping of references to queue, and (iv) add the transaction to the queue with the highest likelihood of conflict.

A {\em policy} consists of a set of design choices. These choices are (i) the form of references that are extracted from the transaction and stored in the History and State components (ii) rules about the assignment of transactions to queues in the Schedule component. In our evaluation, we run experiments with different policies to understand the consequences of various design choices. We introduce the policies explored in this paper with a running example.

\subsection{Count and Fraction policies}
%% count / sum / literal / single / balance
\begin{table}[h]
\begin{center}
\begin{tabular}{llr}
\# & current transactions \\ \hline
1 & \verb|UPDATE stock SET s_quantity=7| \\
  & \verb|WHERE s_i_id=1 AND s_w_id=5| \\
2 & \verb|UPDATE stock SET s_quantity=7| \\
  & \verb|WHERE s_i_id=1 AND s_w_id=5| \\
3 & \verb|UPDATE stock SET s_quantity=6| \\
  & \verb|WHERE s_i_id=2 AND s_w_id=5| \\
4 & \verb|UPDATE stock SET s_quantity=6| \\
  & \verb|WHERE s_i_id=2 AND s_w_id=5| \\
  & \\ % make some space between tables
\end{tabular}

\begin{tabular}{c|c|c}
\multicolumn{3}{c}{\textbf{History}} \\
reference           & abort & commit \\ \hline
\verb|s_quantity=6| & 20     & 60 \\
\verb|s_quantity=7| & 40     & 20 \\
\verb|s_i_id=1|     & 20     & 20 \\
\verb|s_i_id=2|     & 0     & 20 \\
\verb|s_w_id=5|     & 20     & 20 \\
\multicolumn{3}{c}{ } \\ % make some space between tables
\end{tabular}
\begin{tabular}{c|c|c|c|c|c|c}
\multicolumn{7}{c}{\textbf{State}} \\
                    & \multicolumn{3}{c}{queue} & \multicolumn{3}{c}{} \\
reference           & 1 & 2 & 3    & R   & T & Q             \\ \hline
\verb|s_i_id=1|     & 2 & 0 & 0    & 3.1 & 2 & 1              \\
\verb|s_w_id=5|     & 2 & 1 & 1    & 3.0 & 4 & 1           \\

\verb|s_i_id=2|     & 0 & 1 & 1    & 1.9 & 2 & 2              \\ 
\verb|s_quantity=7| & 2 & 1 & 1    & 1.9 & 2 & 1              \\
\verb|s_quantity=6| & 0 & 1 & 1    & 1.9 & 2 & 2              \\ \hline
total               & 6 & 4 & 4    & \multicolumn{3}{c}{} \\

\end{tabular}
\caption{\textbf{Example 1} -- An example of an input sequence, History and State for the repeated execution of the SQL in \cref{fig:sql}. Column `reference' is the extracted representations of the transaction's statements. Column `abort' is the historical count of the number of transactions containing a reference that have aborted. Column `commit' is the count of the number of commits in the same way. A transaction may abort several times before committing. The `queue' columns are the historical count of the transactions scheduled into the reference queue. Row `total' is the total of the queue columns. The R column is the arrival rate of transactions with the associated reference. The T column is the total transactions for this reference in the system. The Q column is the queue that contains the largest number of references. }
\label{tab:example}
\end{center}
\end{table}

\textbf{Example 1}:
Suppose the system is in the state described in \cref{tab:example}. Consider the arrival of a new transaction (\cref{fig:sql}) where \verb|$1=6|, \verb|$2=2|, and \verb|$3=5|.  
The system extracts the references \verb|s_quantity=6|, \verb|s_i_id=2| and \verb|s_d_id=5|. It then looks up the references in the History, and finds the number of aborts as 20, 0 and 20, respectively. 
The system then scores the likelihood of each queue containing a transaction that will conflict with the arriving transaction by summing the product of the abort count with the reference queue count in the State table. For queue one: $20\cdot 0 + 0\cdot 0+2\cdot 2=24$, queue two: $20\cdot 1+0\cdot 1+2\cdot 1=22$, and queue three: $20\cdot 1 + 0\cdot 1+2\cdot 1=22$. The system then chooses the queue with the largest score (queue one), assigns the transaction in the wining queue, and then increments the state values to reflect this assigned transaction.
An entire transaction is always placed onto a single run queue.

Consider the subsequent arrival of a new transaction that contains the reference \verb|s_i_id=3|. The above score cannot be computed because the reference does not occur in the History.
In this case, a {\em default} strategy is used: chose the queue with the smallest total number of references.
%% \\ \vspace{-0.05in}

In Example 1, the strength of evidence for a reference to indicate a transaction abort is the historical count of the number of aborts. We term this design decision the {\em Count} policy.
Intuitively, however, another metric could be used: the fraction of aborts over the total attempted (aborted + committed) transactions for a reference. We term this design decision the {\em Fraction} policy.

\textbf{Example 2}: Rerunning Example 1 with this policy, the same references \verb|s_quantity=6|, \verb|s_i_id=2| and \verb|s_d_id=5| are extracted and looked up in the History table, but now, instead of counts, the fractions $20/(20+60)=0.25$, $0/(0+20)=0.0$ and $20/(20+20)=0.5$ are computed. The system then scores the likelihood of an abort for a queue by multiplying the fractions by the state reference values, and then adding. For queue one: $0.25\cdot 0 + 0.0\cdot 0 + 0.33\cdot 2 = 0.66$, queue two: $0.25\cdot 1 + 0.0\cdot 1 + 0.33\cdot 1 = 0.58$, and queue three: $0.33\cdot 1 + 0.0\cdot 1+0.50\cdot 1 = 0.83$. The system then chooses the queue with the largest score (queue three). 

Compared to the count policy, the fraction score is a localized one. 
It considers each reference independently of the total workload, in this sense: one reference may have a higher abort rate than another, regardless of the frequency that the reference appears in the workload.
In contrast, the count policy is a global one. 
It measures the number of aborts of transactions that contain a reference and thus (indirectly) measures the total work lost, with the assumption that every abort costs the same. This lost work is charged to the references that appear in the transaction. 
Our first experiment compares the fraction and count policies.
 
 %% ==================================================================
%% Max
%% ==================================================================
\subsection{Sum and Max policies}

% describe count/max/literal/single/balance

% show tpc-c occ/2pl results from count/max/literal/single/balance

% and count/max/canonical/single/balance

% same three graphs as above

%describe why literal is problematic
Revising the basic steps of determining a score for a transaction, consider the step of the score computation that sums across all the references in a transaction. The intuition behind this step is that transactions should weight the evidence from all the references that occur. We term this design decision the {\em Sum} policy. However, a different design choice at this point is to focus on a single reference ``hot spot'' that historically has the highest abort count across the workload and use only this reference in the score the transaction with the state table. This design decision is called the {\em Max} policy.
\\ \vspace{-0.05in}

\textbf{Example 3}: Considering again the state of the system from Example 1 (\cref{tab:example}). A new transactions arrives and references \verb|s_quantity=6|, \verb|s_i_id=2| and \verb|s_d_id=5| are extracted and (in the Count policy) the abort counts \verb|20|, \verb|2| and \verb|20| are fetched from the History component.  For the Max policy, the reference with the maximum score is retained and the rest are discarded, leaving \verb|s_w_id=5| (ties are broken arbitrarily). Then this reference is used to compute the scores in the usual manner: queue one $20\cdot 2 = 40$, queue two $20\cdot 1 = 20$, and queue three $20\cdot 1 = 20$. The queue with the maximum count is chosen (queue one).

Comparing these two policies examines two opposing views of the evidence provided by a reference. The Max policy attempts to find the single best reference to use to group transactions into queues, somewhat like a data partition policy. 
In the Sum policy, all references are equally weighted to determine a particular queue.

Note that the Count/Fraction policies are orthogonal to the Sum/Max policies. In Example 3, the Fraction/Max policy for references \verb|s_quantity=6|, \verb|s_i_id=2| and \verb|s_d_id=5| extracts and looks up fractions the History component, returning the fractions $0.25$, $0.0$ and $0.5$, respectively. The reference with the maximum score is retain (\verb|s_d_id=5|). Then the scores for each queue are computed: queue one $0.25\cdot 2 = 0.50$, queue two $0.25\cdot 1 = 0.25$, and queue three $0.25\cdot 1 = 0.25$. The queue with the maximum score is chosen (queue one).
 
\subsection{Literal and Canonical policies}

For all policies so far references are literally extracted from each transaction. Those literal references are used by the History and State components. This design decision we call the {\em Literal} policy. 

However, database administrators partition databases based on an underlying domain concept such as warehouse id. This concept is not captured by literal references because, e.g., foreign key references to the same underlying domain have different literal references. In this section we consider a canonical form of references where references are normalized to the underlying domain. 
For example, for the SQL of~\cref{fig:sql} with parameters \verb|$1=6|, \verb|$2=2|, and \verb|$3=5|, the {\em Canonical} policy refers the underlying domain, thus the canonical references are \verb|s_quantity=6|, \verb|i_id=2| and \verb|w_id=5|.

\begin{table}[h]
\begin{center}
\begin{tabular}{llr}
\# & current transactions \\ \hline
1 & \verb|UPDATE stock SET s_quantity=7| \\
  & \verb|WHERE s_i_id=2 AND s_w_id=5| \\
2 & \verb|UPDATE customer SET c_w_id=6| \\
  & \verb|WHERE c_c_id=11| \\
3 & \verb|INSERT INTO new_orders VALUES| \\
  & \verb|(no_o_id=10, no_c_id=11, no_w_id=5)| \\
  & \\ % make some space between tables
\end{tabular}

\begin{tabular}{c|c|c}
\multicolumn{3}{c}{\textbf{History}} \\
reference       & abort  & commit \\ \hline
\verb|i_id=2|   & 10     & 10 \\
\verb|o_id=10|  & 10     & 10 \\
\verb|c_id=11|  & 10     & 10 \\
\verb|w_id=5|   & 30     & 10 \\
\verb|s_quantity=7| & 20  & 20 \\
\multicolumn{3}{c}{ } \\ % make some space between tables
\end{tabular}
\begin{tabular}{c|c|c|c|c|c|c}
\multicolumn{7}{c}{\textbf{State}} \\
                 & \multicolumn{3}{c}{queue}  & \multicolumn{3}{c}{} \\
reference           & 1 & 2 & 3  & R   & T & Q               \\ \hline
\verb|i_id=2|       & 1 & 0 & 0  & 1.1 & 1 & 1              \\
\verb|w_id=5|       & 1 & 1 & 1  & 4.1 & 3 & 1               \\
\verb|c_id=11|      & 0 & 1 & 1  & 3.1 & 2 & 2              \\ 
\verb|s_quantity=7| & 1 & 0 & 0  & 0.0 & 1 & 1              \\ 
\verb|o_id=10|      & 0 & 0 & 1  & 0.0 & 1 & 3              \\ \hline
total               & 3 & 2 & 3  & \multicolumn{3}{c}{}\\

\end{tabular}
\caption{\textbf{Example 2} -- An example of an input sequence, History and State. Column `reference' is the canonical extracted representations of the transaction's statements. Column `abort' is the historical count of the number of transactions containing a reference that have aborted. Column `commit' is the count of the number of commits in the same way. A transaction may abort several times before committing. The `queue' columns are the historical count of the transactions with a reference queued. Row `total' is the total of the queue columns. The R column is the arrival rate of transactions with the associated reference. The T column is the total number of transactions with this reference in the system. The Q column is the queue that contains the largest number of references.}
\label{tab:example2}
\end{center}
\end{table}

\textbf{Example 4}: Consider the Fraction/Max/Canonical policy for the data in \cref{tab:example2}. Suppose transaction from \cref{fig:sql} arrives where \verb|$1=6|, \verb|$2=2|, and \verb|$3=5|.
Step 1 the canonical policy would extract references \verb|s_quantity=6|, \verb|i_id=2| and \verb|w_id=5| from the transaction. Step 2 would look up the fractions $0.5$, $0.5$, and $0.75$ for the references in the History table. Step 3 would chose the reference \verb|w_id=5| with the maximum fraction. Step 4 computes the scores for each queue: queue one $0.75$, queue two $0.75$, and queue three $0.75$. Picking the queue with the highest score results in a three way tie.
In this case the default policy is used and the queue with the smallest number of references is chosen, so the transaction is assigned to queue two.

This policy comparison of Literal versus Canonical explores two effects.
First, the literal policy dimension results in a more fine-grained and precise representation, but also a sparser representation because there are many more unique literal references than canonical ones.
Canonical references produces a smaller, more compact, less fine-grained representation.
Second, the canonical policy requires additional implementation work.
Literal references are straight forward to implement with a string copy of an element of the parse tree. 
Implementing canonicalization requires deeper analysis to trace each reference to its underlying domain.
The information required for the trace may be explicitly recorded in the schema (through, say, explicit domain statements) or a model can be constructed to determine likely candidates~\cite{Bowman:2005}.
For our experiments, we simply hand-code the canonical references for a transaction.

\subsection{Single and All policies}

In addition, we compared the extraction of single literals as references (the Single policy) versus using the entire condition as a reference (the All policy). For example, in \cref{tab:example2}, the first transaction listed has a boolean AND condition in the WHERE clause. With the All policy, this clause generates a single (literal) reference \verb|s_i_id=2 AND s_w_id=5| instead of two references. With the canonical and all policies, the single reference is \verb|i_id=2 AND w_id=5|.
The Single policy slices up complex conditionals and models each reference individually. This technique reduces the total number of references but does not model the boolean condition in the conditional at all. The All policy assumes essentially that a condition acts as a partial 'key' that identifies a transaction, since conditions are much more likely to be unique.

%% ==================================================================
%% Balance vs. random
%% ==================================================================
%\subsection{Default policy - Balance and Random}

%%%%%%% Below is what I explained about balance and random %%%%%%
%Finally, for our discussion of policy design considerations, we consider the default policy. As described initially, the default policy is applied when the Schedule component cannot find references either in the History or State, or when the scoring step produces a tie. This situation typically arises when the system starts running (and the History table is already populated). In this case, the queue with the least total references is chosen. We name this case the {\em Balanced} policy. 
%The alternative, more in keeping with traditional transaction processing, is to randomly assign the transaction to a queue. We name this case the {\em Random} policy.
 
%% ==================================================================
%% Experimental Framework
%% ==================================================================
\section{Experimental Framework}
\label{sec:framework}

The experimental framework focuses on isolating external factors from the system to highlight any differences in the policies.
Experiments are run on a server that is otherwise idle.
%\textbf{The server used for the experiment is a Dell PowerEdge R320 with 1 CPU, 6 hardware threads and 32 gigabytes of memory. We turn on hyper-threading for a total of 12 threads.} 
The server used for the experiment contains 20 cores from Intel(R) Xeon CPU E5-2680 v2 @ 2.80GHz with 132 GB of memory.
One thread is pinned to workload generation and Policy computations exclusively. 
The remaining 19 worker threads execute transactions and the History and State component code that maintains the data.
This arrangement separates out the cost of workload generation from our measurements, since the worker threads never wait for the workload generation thread.

All experiments are run on the \peloton system~\cite{Pavlo:2017-real}, an open source main-memory database system.
As part of the experiment framework implementation, for each transaction benchmark, we manually label references in the transaction workload for the  literal and canonical rules. 
The steps in an experiment are as follows.

\squishitemize
    \item \textbf{Phase I - History Initialization}
    \squishlist
        \item Initialize the random number generator to a new unique seed.
        \item Initialize and start the database for the benchmark.
        \item Generate a large batch of transactions.
        \item Execute transactions for 5 seconds under a given policy, but schedule transactions by random assignment to a queue instead of by the Schedule component. Populate the History component according to the reference sub-policy, but do not populate the State component.
    \squishend
    
    \item \textbf{Phase II - Execution}
    \squishlist
        \item Initialize and start the database for the benchmark.
        \item Generate a large batch of transactions.
        \item Turn on measurements, then schedule and execute transactions for 180 seconds under a given policy, using the previous History component as read-only. Populate the State component in real time. 
        \item Stop the database.
    \squishend

    \item Repeat the above two phases 10 times.
    \item Average the measurements across the 10 experiment runs.
\squishend

Note that the experimental framework separates out in Phase I the initial population of the History from the execution of transactions in Phase II. This technique isolates the policy behavior from the gathering of accurate statistics about the workload. 
In addition, the use of random scheduling during this phase insures that the History statistics are an accurate reflection of the underlying workload. 
This structure of the experiment points an underlying feedback loop of any particular policy - the policy itself affects the statistics stored, which in turn affect the execution of the policy.
Without careful management, this feedback loop can lead to instabilities over time.

Response time is measured for a transaction from the start of the policy generation code until the transaction commits. 
If a transaction aborts, it is immediately restarted, and this time is included in the response time recorded for the transaction.
Throughput is measured as the total number of transactions to commit divided by experiment time.
The abort rate is measured as the total number of aborts divided by the sum of the total number of aborts and commits.

We use three popular benchmarks, TPC-C~\cite{TPCC}, SmallBank~\cite{Cahill:2008:SIS:1376616.1376690} and TATP~\cite{TATP}, to evaluate the proposed methods. The benchmark parameters are listed in \cref{tab:parameters} and the workloads are listed in \cref{tab:workload}. We added a hot spot for value references to SmallBank and TATP to increase the frequency of aborts while remaining true to the realism of the benchmarks.

%% ==================================================================
%% Results and Discussions
%% ==================================================================
\section{Results and Discussions}

The design space we explore contains multiple independent dimensions, so we ran a k-factor experiment to determine baseline performance compared to two other systems (\cref{tab:tpc-many}).  One other system is a vanilla policy with random assignment of transactions to threads (labeled 'Original' in the figure). This third result serves as the baseline for an existing system. In addition, we compare to a Hard Partition assignment of transactions to threads based on the warehouse id in the transaction. This result serves as the best case performance to match. Experimentally, running an unthrottled system, we step-wise increase the number of threads and track the corresponding increase in abort rate and throughput. We performed this experiment both for OCC and 2PL.

\subsection{Literal and Canonical Policies}

\begin{figure*}[h]
    \subfigure[Abort rate]{
    \label{fig:abort-rate-literal-canonical-occ} %% label 
        \begin{minipage}[b]{0.5\textwidth}
            \centering
            \includegraphics[width=3in]{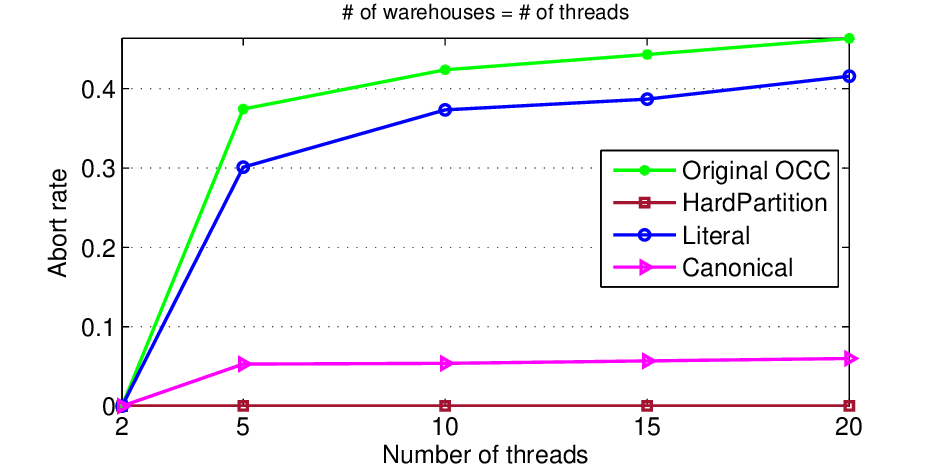}
        \end{minipage}
    }%
    \subfigure[Throughput]{
    \label{fig:throughput-literal-canonical-occ} %% label 
        \begin{minipage}[b]{0.5\textwidth}
            \centering
            \includegraphics[width=3in]{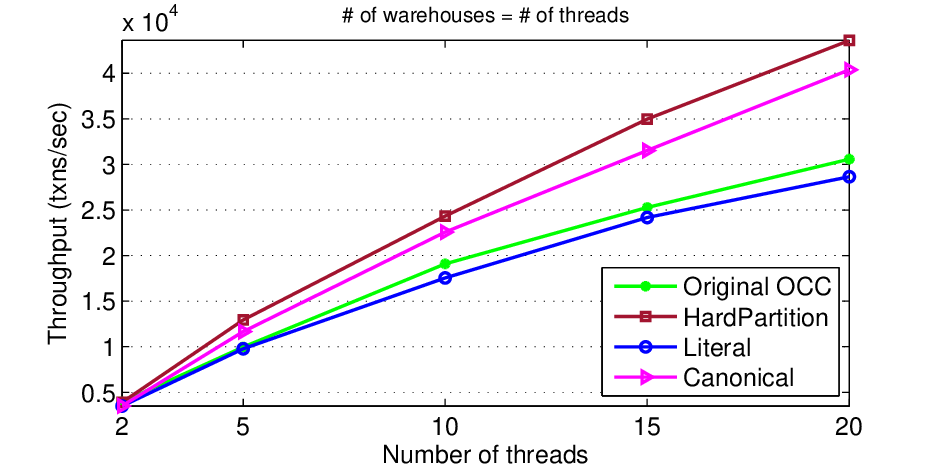}
        \end{minipage}
    }
    % \subfigure[Response time]{
    % \label{fig:response-time-literal-canonical-occ} %% label 
    %     \begin{minipage}[b]{0.32\textwidth}
    %         \centering
    %         \includegraphics[width=2.1in]{fig/response_time_literal_canonical_tpcc_occ.eps}
    %     \end{minipage}
    % }
\caption{\textbf{OCC Throughput of Literal and Canonical} -- The throughput performance of TPC-C over OCC comparing Literal with Canonical using Count/Max/Single. Graph (a) shows Canonical can decrease the abort rate from 45\% to 6\%. Accordingly, as shown in Graph (b) the throughput of Canonical is close to Hard Partition and is $\sim$30\% higher to Original OCC when the number of threads is 20.}
\label{fig:literal-canonical-occ} %% label for entire figure
\end{figure*}

\begin{figure*}
    \subfigure[Abort rate]{
    \label{fig:abort-rate-literal-canonical-2pl} %% label 
        \begin{minipage}[b]{0.5\textwidth}
            \centering
            \includegraphics[width=3in]{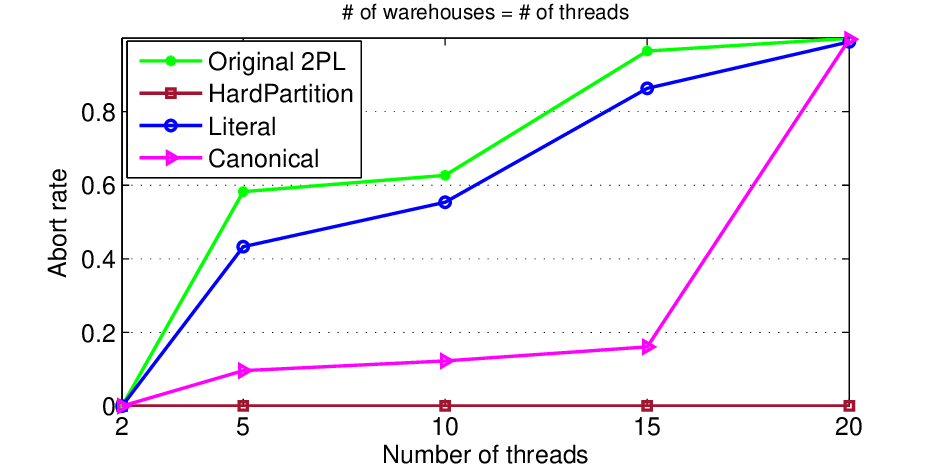}
        \end{minipage}
    }%
    \subfigure[Throughput]{
    \label{fig:throughput-literal-canonical-2pl} %% label 
        \begin{minipage}[b]{0.5\textwidth}
            \centering
            \includegraphics[width=3in]{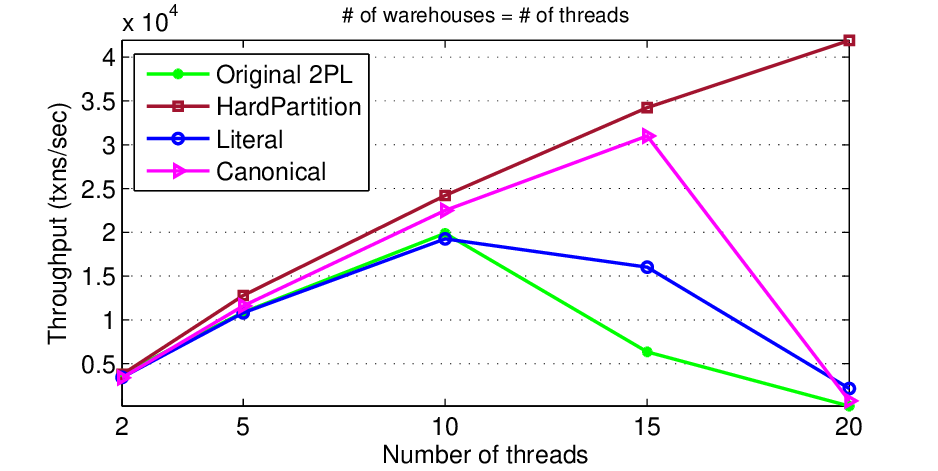}
        \end{minipage}
    }
    % \subfigure[Response time]{
    % \label{fig:response-time-literal-canonical-2pl} %% label 
    %     \begin{minipage}[b]{0.32\textwidth}
    %         \centering
    %         \includegraphics[width=2.1in]{fig/response_time_literal_canonical_tpcc_2pl.eps}
    %     \end{minipage}
    % }
\caption{\textbf{2PL Throughput of Literal and Canonical} -- The performance of TPC-C over 2PL comparing Literal with Canonical using Count/Max/Single. Lock contention occurs in Original 2PL, Literal and Canonical. But Canonical is better than both Original 2PL and Literal. }
\label{fig:literal-canonical-2pl} %% label for entire figure
\end{figure*}

To determine more detail, we ran an experiment that compares the Literal and Canonical policies with (Count, Max, Single) policies held constant on TPC-C. Fixing the number of warehouses equal to the number of threads, we increased the threads to a maximum of 20. The results for OCC (\cref{fig:literal-canonical-occ}) show that Canonical performs well on OCC closely matching the performance of Hard Partition. The Literal policy offers only a modest improvement on OCC. Canonical successfully (appoximately) groups by warehouse id. Note however, that under 2PL (\cref{fig:literal-canonical-2pl}), the Canonical policy performs well until very high contention at 20 threads.

%We ran an experiment that compares the Literal and Canonical policies, using Count and Max, over TPC-C on the OCC protocol (\cref{fig:literal-canonical-occ}). The results are dramatic, leading to significantly lower abort rate, higher throughput and reduced response times. In fact, in unthrottled experiment, Canonical improves throughput by $\sim$40\% comparing to Original OCC. Under the throttled experiment, Canonical matches Hard Partition performance up to 14,000 transactions per second before falling behind. Similar trends can be observed under 2PL (\cref{fig:literal-canonical-occ}).

%%%%%%%%%%% count-max-literal-canonical-occ %%%%%%%%%%
%%%%%%%%%%%%%%%%%%%%%%%%%%%%%%%%%%%%

%% ==================================================================
%% All
%% ==================================================================
%\subsection{All}

%investigate if single is a good idea, show a table? to show that it is

%%%%%%%%%%% count-max-literal-canonical-2pl %%%%%%%%%%
%%%%%%%%%%%%%%%%%%%%%%%%%%%%%%%%%%%%

\begin{figure}[h]
\centering
\includegraphics[width=8.9cm]{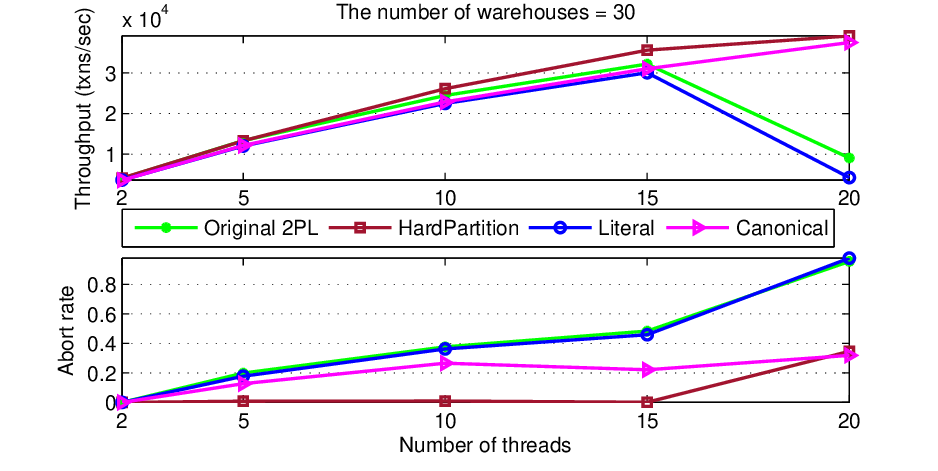}
\caption{\textbf{2PL Throughput and Abort Rate with 30 Warehouses} --The performance of TPC-C over 2PL with 30 warehouses. In this case, thrashing still occurs in Original 2PL and Literal. But Canonical doesn't have the thrashing issue, because 30 warehouses brings a lower abort rate.  }
\label{fig:literal-canonical-2pl-30w} %% label for entire figure
\end{figure}

%%%% exlain for 2pl-30warehouses %%%
To test the hypothesis that high contention is the cause of poor performance for the best policy under 2PL, we ran an experiment that fixed the number of warehouses at 30 and varied the number of threads (\cref{fig:literal-canonical-2pl-30w}). For Hard Partition the abort rate increases to 35\% when the number of threads is 20. This behavior results from the following effect. The workload is not balance (20 threads and 30 warehouses) so some threads become idle. Idle threads  steal transactions from other threads, which under hard partition are guaranteed to refer to a warehouse id of another thread. So the abort rate increases for Hard Partition. For Canonical, the abort rate increases as well, but not so sharply as Hard Partition. For the remainder of the Results section, we run OCC with 19 warehouses and 19 worker threads, but for 2PL we run 30 warehouses and 19 worker threads.

%%%%%%%%%%% count-max-literal-canonical-2pl-30warehouses %%%%%%%%%%

\begin{figure}[t]
\centering
\includegraphics[width=8.9cm]{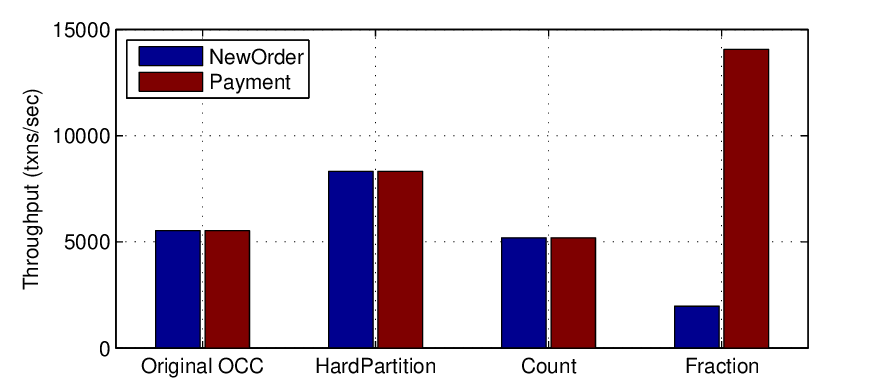}
\caption{A comparison of the throughput of NewOrder versus Payment. The numbers of NewOrder and Payment are not equal in Fraction, which we term the workload distortion problem.}
\label{fig:balance-policy}
\end{figure}
% %%%%%%%%%%%%%%%%%%%%%%%%%%%%%%%%%%%%

%%%%%%%%%%%%%%%%%%%%%%%%%%%%%%%%%%%%%%%%%%%%%%%%%%%%%%%%%%%%%%%%%%%%

We examined the logs of the system and detected an anomalous result for Fraction that is the result of two interacting effects. The first effect involves the scoring of Fraction.  We examined the log of transaction assignments in the Schedule component and observed that the Fraction policy sorts NewOrder and Payment transactions into separate queues. NewOrder takes much longer to execute than Payment, so once transactions are sorted into separate queues, the Payment queues execute much, much more quickly. Each time a Payment order finishes, the workload generates a new NewOrder or Payment transaction with even probability. If the result is a NewOrder transaction, it is simply queued waiting behind the currently running NewOrder. If the result is a Payment transaction, it is queued and then quickly executed. This behavior results in many more Payment than NewOrder transactions finishing (\cref{fig:balance-policy}).
Of course, this workload distortion behavior is a side-effect of our experimental workload design interacting with the sorting behavior of the Fraction policy. The workload we observe would not occur in the real world where NewOrder and Payment transactions are generated roughly at the same rate. The behavior does not occur with the count policy, which uses warehouse id as the main variable to queue both NewOrder and Payment.

%% ==================================================================
%% Other Combinations
%% ==================================================================
\subsection{$2^k$ Factor Experiment}

We ran all possible combinations of the policies under OCC and 2PL for TPC-C to compare policies. In this experiment, we throttled the system by fixing average response time to 500 ms and measuring throughput and abort rate at this response time. This experimental design roughly corresponds to the client response time limit in TPC-C. 
The results in~\cref{tab:tpc-many} show that generally. But the best performing policy is Count Max Canonical Single.

\begin{table}[h]
\begin{center}
\begin{tabular}{l|l|l|l|c|c|c|c}
  \multicolumn{4}{l}{\multirow{2}{*}{{\bf Policy}}} & \multicolumn{2}{c}{\bf TPCC-OCC} & \multicolumn{2}{c}{\bf TPCC-2PL} \\ 
  \multicolumn{4}{l|}{} & tps & rate & tps & rate \\ \cline{1-8}
  \multicolumn{4}{l|}{Original} & 35.567 & 0.45 & 1,412 & 0.99 \\ \cline{1-8}
  \multicolumn{4}{l|}{Hard Partition} & 43,137 & 0.01 & 39,338 & 0.29 \\ \cline{1-8}
  \multirow{8}{*}{\rotatebox[origin=c]{90}{Fraction}} & \multirow{4}{*}{\rotatebox[origin=c]{90}{Sum}} & \multirow{2}{*}{Literal} & Single & 29,863 & 0.45 & 31,776 & 0.69 \\ \cline{4-8}
  & & & All & 28,670 & 0.44 & 36,146 & 0.51  \\ \cline{3-8}
  & & \multirow{2}{*}{Canonical} & Single & 29,265 & 0.44 & 31,715 & 0.66 \\ \cline{4-8}
  & & & All  & 29,489 & 0.43 & 30,154 & 0.71 \\ \cline{2-8}
  & \multirow{4}{*}{\rotatebox[origin=c]{90}{Max}} & \multirow{2}{*}{Literal} & Single  & 29,399 & 0.44 & 37,195 & 0.52 \\ \cline{4-8}
  & & & All & 28,659 & 0.45 & 4,986 & 0.97 \\ \cline{3-8}
  & & \multirow{2}{*}{Canonical} & Single & 29,489 & 0.44 & 29,655 & 0.72 \\ \cline{4-8}
  & & & All & 29,463 & 0.43 & 30,289 & 0.72 \\ \cline{1-8}
  \multirow{8}{*}{\rotatebox[origin=c]{90}{Count}} & \multirow{4}{*}{\rotatebox[origin=c]{90}{Sum}} & \multirow{2}{*}{Literal} & Single & 29,587 & 0.44 & 16,899 & 0.89 \\ \cline{4-8}
  & & & All & 28,865 & 0.44 & 19,998 & 0.86 \\ \cline{3-8}
  & & \multirow{2}{*}{Canonical} & Single & 29,783 & 0.44 & 6,748 & 0.97 \\ \cline{4-8}
  & & & All & 28,990 & 0.44 & 16,787 & 0.91 \\ \cline{2-8}
  & \multirow{4}{*}{\rotatebox[origin=c]{90}{Max}} & \multirow{2}{*}{Literal} & Single & 27,899 & 0.41 & 28,881 & 0.71 \\ \cline{4-8}
  & & & All & 32,175 & 0.29 & 36,578 & 0.38 \\ \cline{3-8}
  & & \multirow{2}{*}{Canonical} & Single & {\bf 40,612} & {\bf 0.07} & {\bf 38,125} & {\bf 0.29} \\ \cline{4-8}
  & & & All & 33,763 & 0.24 & 31,590 & 0.71 \\
  % \cline{1-8}
\end{tabular}
\caption{\textbf{$2^k$ Factor Experiment Results} -- The performance of an enumeration of policy combinations for TPC-C.  Original means that transactions are simply randomly scheduled into queues as per usual. The column `tps' reports transactions per second and column `rate' reports abort rate. The OCC is under 19 warehouses and 2PL is under 30 warehouses. }
\label{tab:tpc-many}
\end{center}
\end{table}

\subsection{SmallBank and TATP}

%%%%%%%% smallbank-occ %%%%%%%%%%
\begin{figure*}[h]
    \subfigure[Abort rate]{
    \label{fig:abort-rate-smallbank-occ} %% label 
        \begin{minipage}[b]{0.32\textwidth}
            \centering
            \includegraphics[width=2.1in]{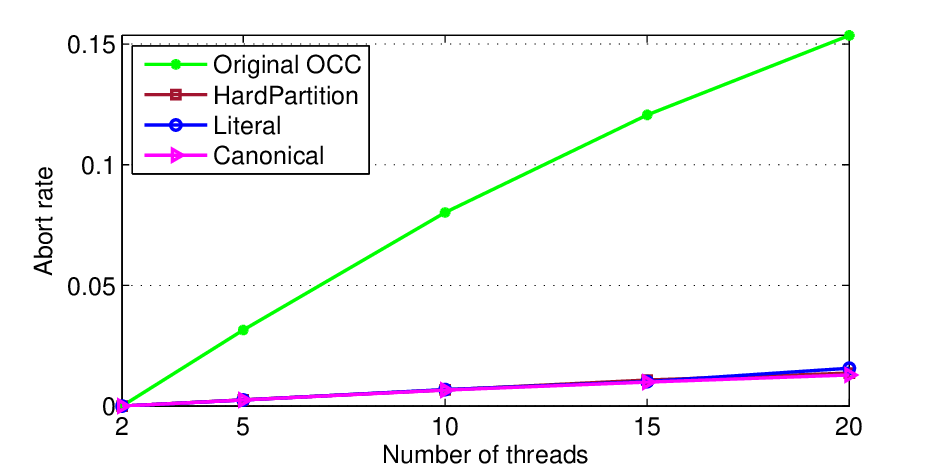}
        \end{minipage}
    }%
    \subfigure[Throughput]{
    \label{fig:throughput-smallbank-occ} %% label 
        \begin{minipage}[b]{0.32\textwidth}
            \centering
            \includegraphics[width=2.1in]{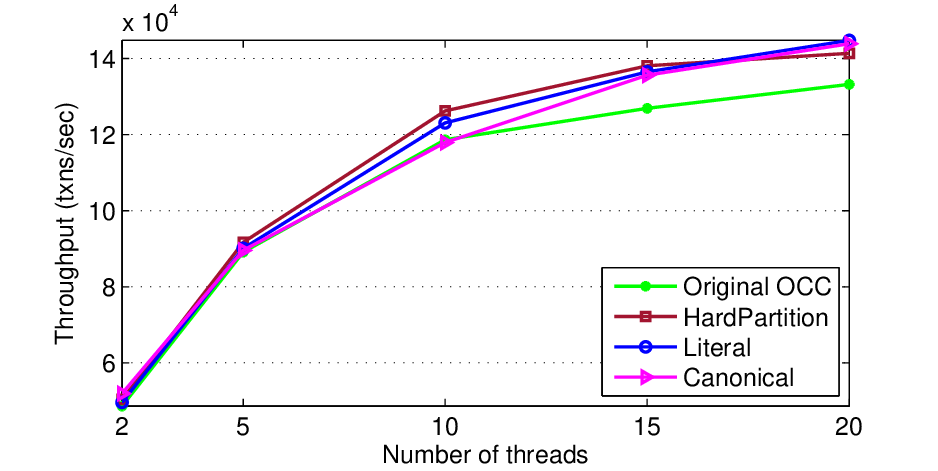}
        \end{minipage}
    }
    \subfigure[Response time]{
    \label{fig:response-time-smallbank-occ} %% label 
        \begin{minipage}[b]{0.32\textwidth}
            \centering
            \includegraphics[width=2.1in]{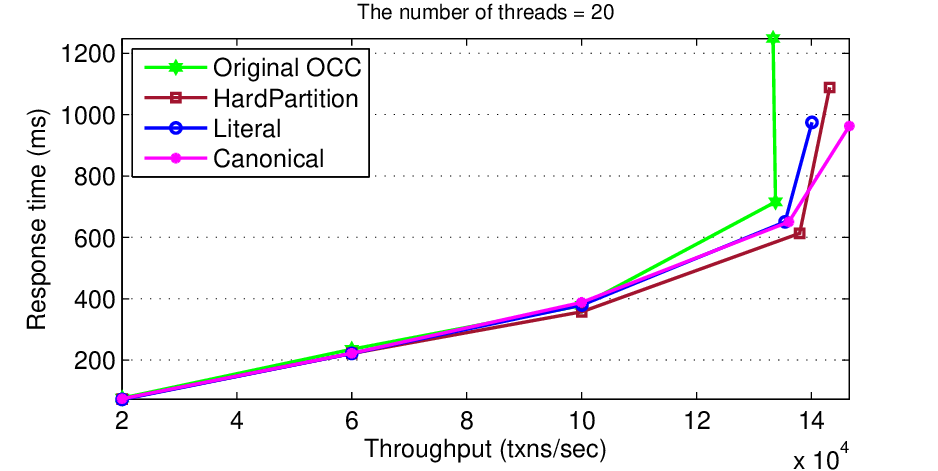}
        \end{minipage}
    }
\caption{\textbf{SmallBank over OCC} -- Graphs (a) and (b) are unthrottled experiments. Graph (c) is a throttled experiment with 20 threads where the arrival rate of transactions is limited to generate a particular throughput value.}
\label{fig:smallbank-occ} %% label for entire figure
\end{figure*}

The best policy discovered in our TPC-C experiments is Count/ Max/~Canonical/~Single, which we name `Canonical' for short. In this section, we examine Canonical's performance over SmallBank and TATP benchmarks. 
%Besides for Canonical, we implement Original OCC/2PL, Hard Partition and the other three Max polices to compare. 

\subsubsection{SmallBank}

In this section we discuss the SmallBank performance of the Canonical policy for both OCC and 2PL.

\textbf{OCC}: \cref{fig:smallbank-occ} presents the results on OCC. Original OCC has an abort rate that increases linearly with additional threads, reaching around 15\% when the number of threads is 20.  The abort rate is about ~2\% for Canonical at this point. The abort rate of Canonical increases slowly as the number of threads are increased.
The corresponding throughput is shown in~\cref{fig:throughput-smallbank-occ}. Compared to Original OCC, the throughput is improved by ~5\% by Canonical when the number of threads is 15. The performance difference is due to most aborts being write-write conflicts (\cref{tab:abort_read_write}). Write-write conflict has a smaller cost to throughput than read-write conflict for OCC, so the throughput is improved less than abort rate.
Canonical has the same abort rate as Hard Partition. In SmallBank, each transaction is based on a customer id. So customer id has the largest conflict compared to other references. %The abort rate induced by 'Amalgamate' transaction is inevitable, because it %includes two customer ids. 
Hard Partition uses customer id to assign transactions to different queues, while Canonical classifies transactions by customer id with largest conflicts. Under higher contention, although Canonical has a higher abort rate than Hard Partition, the gap in throughput is small, around 1\%.

\begin{table}
\begin{center}
\begin{tabular}{lc|r|r|r}
\multicolumn{4}{c}{} \\
Benchmark & Protocol & write abort & read abort & ratio \\ \hline
SmallBank & OCC & 44580 & 27618 & 1.61 \\
SmallBank & 2PL  & 79876 & 19588 & 4.08 \\
TATP      & OCC  & 71556 & 37896 & 1.89 \\
TATP      & 2PL  & 68990 & 11345 & 6.08 \\
TPC-C     & OCC  & 31426 & 18638 & 1.69 \\
TPC-C     & 2PL  & 119889 & 53636 & 2.24 \\
\end{tabular}
\caption{\textbf{The Ratio of Write Abort to Read Abort} -- The measurement of the ratio for all combinations of benchmarks and concurrency control protocols using random assignment of transactions to threads.}
\label{tab:abort_read_write}
\end{center}
\end{table}

Since our policy will compute which queue/thread is the best for a given transaction, this computation time is included for response time. In this experiment, Canonical's response time is slightly higher than Hard Partition and Original OCC when the throughput is less than 120,000 txns/sec, due to this computational cost of assigning the transaction (\cref{fig:response-time-smallbank-occ}). But after throughput is larger than 120,000 txns/sec, the system hits its performance limit and additional transactions simply increase response time while adding very little to throughput.
\\ \vspace{-0.05in}

%%%%%%%% smallbank-2pl %%%%%%%%%%
\begin{figure*}
    \subfigure[Abort rate]{
    \label{fig:abort-rate-smallbank-2pl} %% label 
        \begin{minipage}[b]{0.32\textwidth}
            \centering
            \includegraphics[width=2.1in]{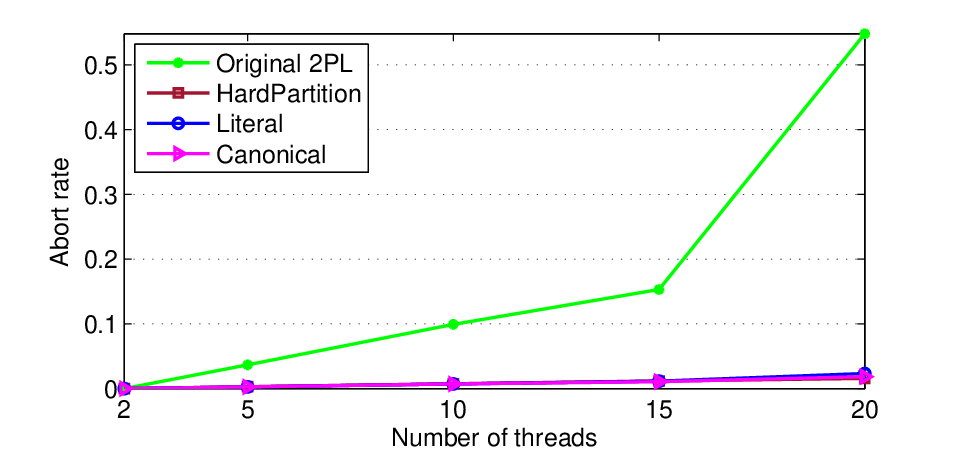}
        \end{minipage}
    }%
    \subfigure[Throughput]{
    \label{fig:throughput-smallbank-2pl} %% label 
        \begin{minipage}[b]{0.32\textwidth}
            \centering
            \includegraphics[width=2.1in]{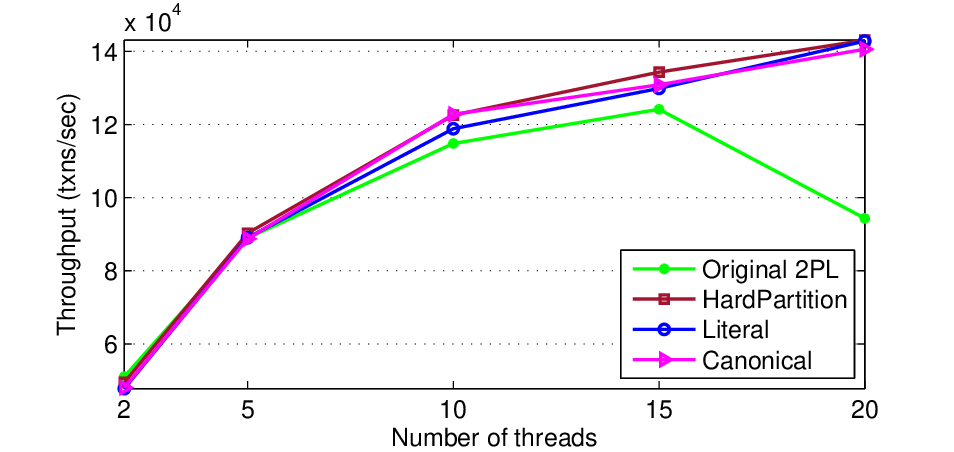}
        \end{minipage}
    }
    \subfigure[Response time]{
    \label{fig:response-time-smallbank-2pl} %% label 
        \begin{minipage}[b]{0.32\textwidth}
            \centering
            \includegraphics[width=2.1in]{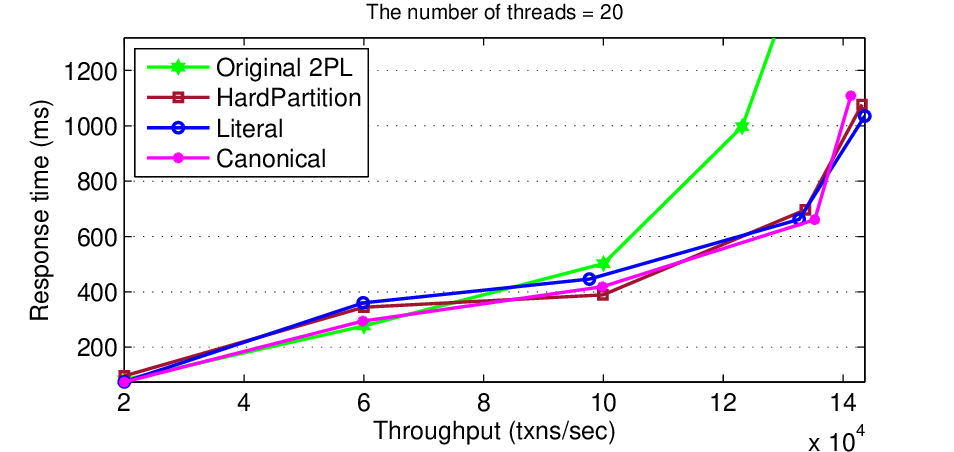}
        \end{minipage}
    }
\caption{\textbf{SmallBank over 2PL} -- The performance of SmallBank over 2PL with Canonical and Literal with Count/Max/Single held constant. Graphs (a) and (b) are unthrottled experiments. Graph (c) is a throttled experiment.}
\label{fig:smallbank-2pl} %% label for entire figure
\end{figure*}

\textbf{2PL}: ~\cref{fig:smallbank-2pl} presents results of 2PL, which is more responsive to transaction abort. Accordingly the abort rate is higher for 2PL than OCC for all algorithms (see~\cref{fig:abort-rate-smallbank-2pl}). Hard Partition performs about as well as Canonical.  For Original 2PL, the abort rate increases linearly until 15 threads where it is $\sim$11\% higher than Canonical, and then increases dramatically at 20 threads. The corresponding throughput is shown in~\cref{fig:throughput-smallbank-2pl}. We can see that Canonical increases throughput and matches Hard Partition. At 20 threads, both policies deliver $\sim$55\% higher throughput.
%There is no validation phase in 2PL, the abort is less expensive for 2PL than OCC. 
For response time (\cref{fig:response-time-smallbank-2pl}), All the policies perform similarly with less than 100,000 TPS. But at this point, contention increases for OCC and response time worsens.  The other policies perform similarly even at 140,000 TPS.

\subsubsection{TATP}
\label{exp:TATP}

\begin{figure*}
    \subfigure[Abort rate]{
    \label{fig:abort-rate-tatp-occ} %% label 
        \begin{minipage}[b]{0.32\textwidth}
            \centering
            \includegraphics[width=2.1in]{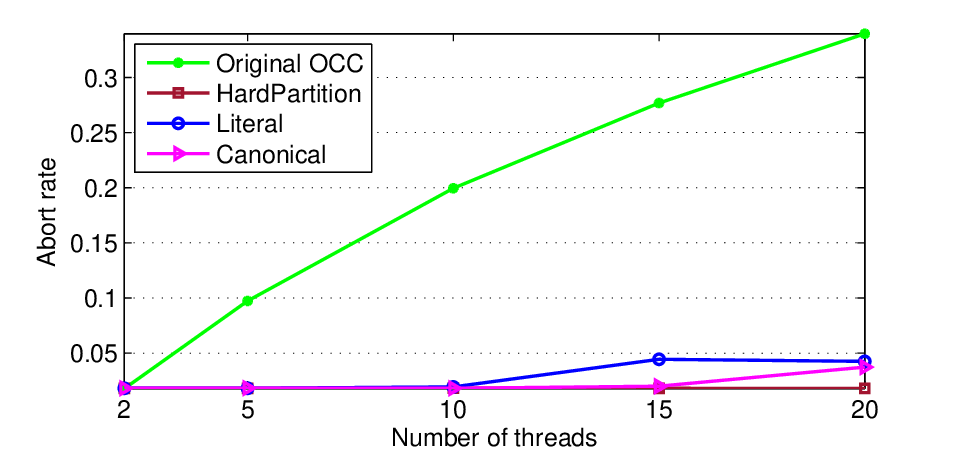}
        \end{minipage}
    }%
    \subfigure[Throughput]{
    \label{fig:throughput-tatp-occ} %% label 
        \begin{minipage}[b]{0.32\textwidth}
            \centering
            \includegraphics[width=2.1in]{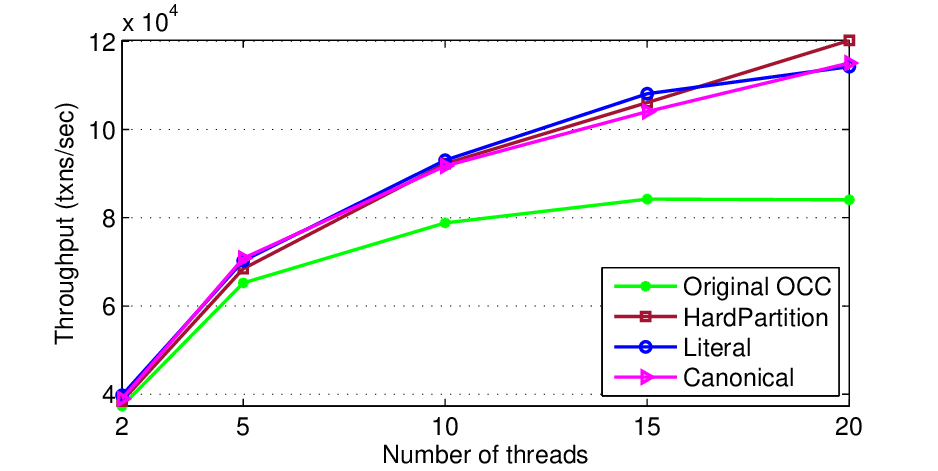}
        \end{minipage}
    }
    \subfigure[Response time]{
    \label{fig:response-time-tatp-occ} %% label 
        \begin{minipage}[b]{0.32\textwidth}
            \centering
            \includegraphics[width=2.1in]{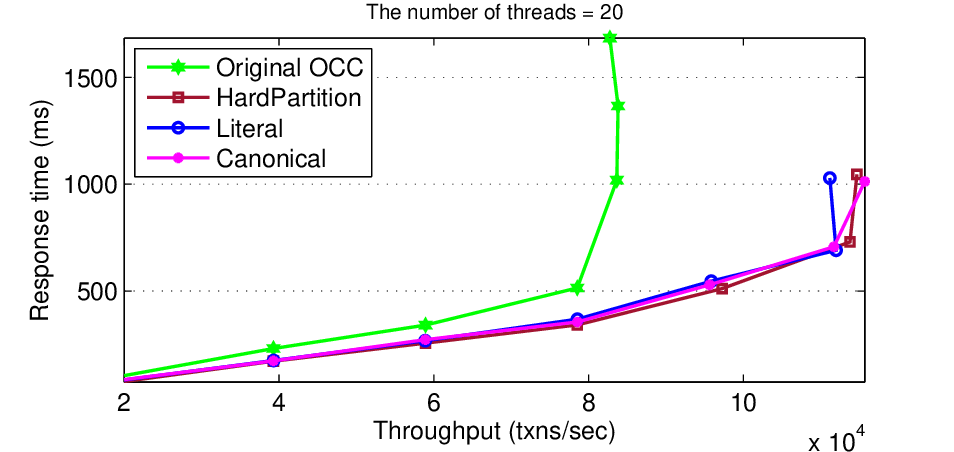}
        \end{minipage}
    }
\caption{\textbf{TATP over OCC} -- The performance of TATP over OCC with Canonical and Literal. Policies Count, Max, and Single are held constant. Graphs (a) and (b) are unthrottled experiments. Graph (c) is a throttled experiment.}
\label{fig:tatp-occ} %% label for entire figure
\end{figure*}

In this section we discuss the TATP performance of the Canonical policy for both OCC and 2PL.
%\\ \vspace{-0.05in}

\textbf{OCC}: \cref{fig:tatp-occ} presents the results of TATP on OCC. TATP is based on a subscriber id for all transactions. Comparing to SmallBank, TATP has more a more transactional structuring including UPDATE, DELETE and INSERT. This structure leads TATP to have a higher abort rate for Original OCC than SmallBank  (~\cref{fig:abort-rate-tatp-occ}). On the other hand, unlike SmallBank, there are no transactions spanning different subscriber ids, so the transactions can easily separated by subscriber id. Therefore, Hard Partition has a flat line for abort rate when the number of threads increases because  Hard Partition uses subscriber id to separate transactions and put them in different queues. At 15 threads the Literal policy abort rate jumps to $\sim$5\% and Canonical has the same jump at 20 threads. The abort rate of Canonical is nearly the same with Hard Partition up to 15 threads, and for both of them, the abort rate is $\sim$1.9\%. This abort rate is induced by INSERT statements in the InsertCallForwarding transaction. Because most tuples already exist in DBMS, the INSERT operation almost always fails unless the tuple doesn't exist. Failed INSERT operations are counted as an aborted transaction, but they are not retried. 
%Given the low abort rate, a correspondingly improved throughput of $\sim$14\% is provided by Canonical compared to Original OCC when the number of threads is 11 (~\cref{fig:throughput-tatp-occ}). 
%There is 4\% gap between throughput (14\%) and abort rate (18\%), which is similar with that (3\%) in SmallBank. For response time (~\cref{fig:response-time-tatp-occ}), unlike SmallBank, Canonical is always better than Original OCC because Original OCC has higher abort rate in TATP than  in SmallBank. 

\label{sec:thrashing}

\begin{figure*}
    \subfigure[Abort rate]{
    \label{fig:abort-rate-tatp-2pl} %% label 
        \begin{minipage}[b]{0.32\textwidth}
            \centering
            \includegraphics[width=2.1in]{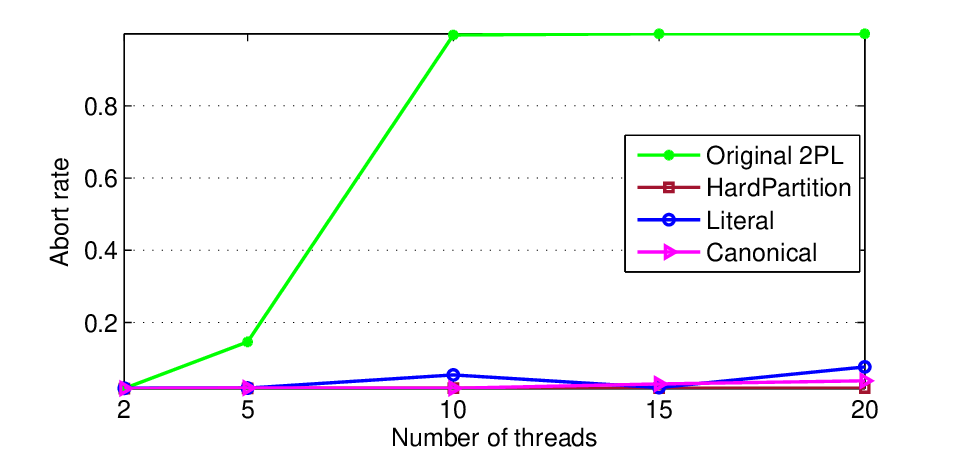}
        \end{minipage}
    }%
    \subfigure[Throughput]{
    \label{fig:throughput-tatp-2pl} %% label 
        \begin{minipage}[b]{0.32\textwidth}
            \centering
            \includegraphics[width=2.1in]{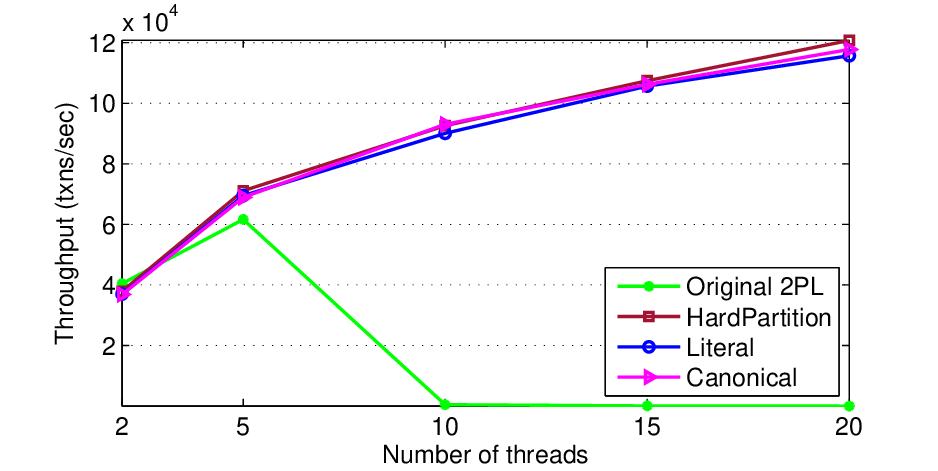}
        \end{minipage}
    }
    \subfigure[Response time]{
    \label{fig:response-time-tatp-2pl} %% label 
        \begin{minipage}[b]{0.32\textwidth}
            \centering
            \includegraphics[width=2.1in]{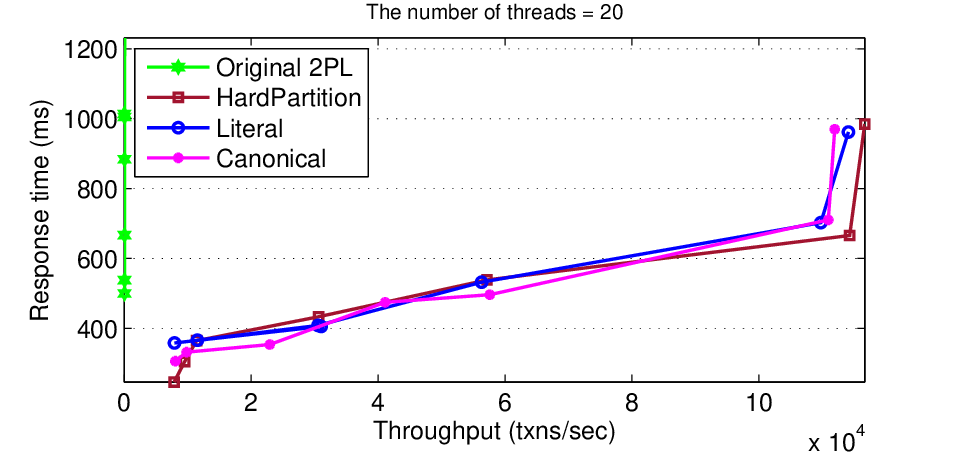}
        \end{minipage}
    }
\caption{\textbf{TATP over 2PL} -- The performance of TATP over 2PL for several polices using Count. Graphs (a) and (b) are unthrottled experiments. Graph (c) is a throttled experiment. The results for Literal and Canonical are almost the same performance because in TATP there is always only a single reference in WHERE clauses. }
\label{fig:tatp-2pl} %% label for entire figure
\end{figure*}

\textbf{2PL and Lock Contention}: \cref{fig:tatp-2pl} shows the results of 2PL running TATP. The figure shows that lock contention~\cite{Yu:2014:SAE:2735508.2735511} occurs when Original 2PL at 10 threads (\cref{fig:abort-rate-tatp-2pl}). The lock contention problem happens when a transaction already holds a lock and requires another lock. This event will cause other transactions to wait, since the lock is held for relatively long time. In TATP under 2PL, transactions are randomly assigned to queues. As the contention rises with more threads,  aborted transactions thrash as they are immediately retried, eventually leading to almost a 100\% abort rate.  However, under the Canonical and Literal policies the this situation is avoided since transactions are partitioned by subscriber id, essentially mimicking Hard Partition.

%%%%%%%% tatp-2pl %%%%%%%%%%
%%%%%%%%%%%%%%%%%%%%%%%%%%%%%%
%% ==================================================================
%% Discussions
%% ==================================================================

%% ==================================================================
%% Continuous Running
%% ==================================================================
\subsection{Continuous Running}
\begin{figure}[h]
\centering
\includegraphics[width=8.9cm]{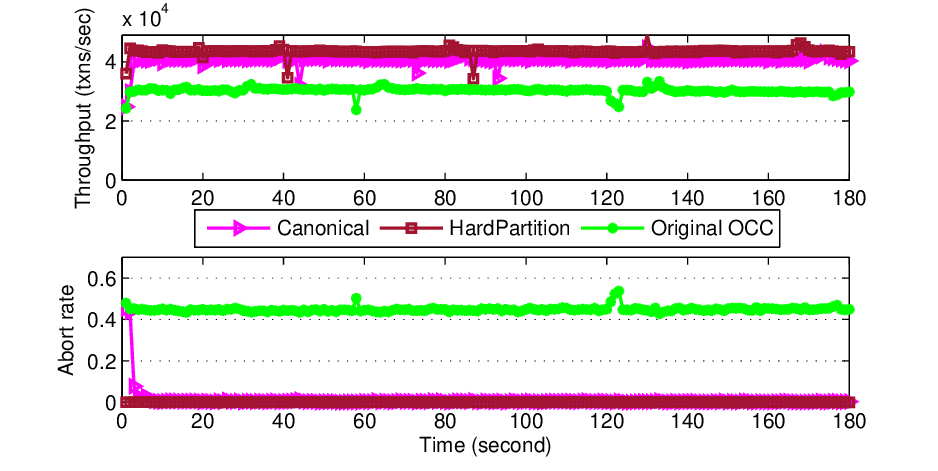}
\caption{The performance with time-continuous in TPC-C. The graphs examine how Canonical performs comparing with HardPartition and Original OCC with 180 seconds running. There are some jitters, because the index in our system uses libcuckoo, and libcuckoo will increase CPU consuming when it resizes.}
\label{fig:continue_thr_abort} %% label for entire figure
\end{figure}

Pls see~\cref{fig:continue_thr_abort}

\begin{figure}[h]
\centering
\includegraphics[width=8.9cm]{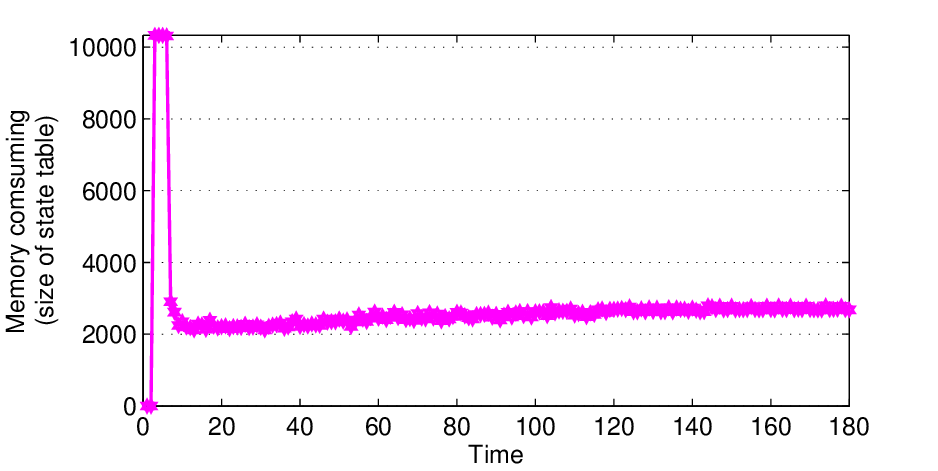}
\caption{Memory consuming of our design (Canonical). There is a spike at the beginning, and after that the memory consuming is flat. The system will wait till the abort rate is table and then deletes the items of State Table, That is why there is a spike for the memory consuming.}
\label{fig:continue_memory} %% label for entire figure
\end{figure}

For memory consuming, pls see~\cref{fig:continue_memory}.

\begin{figure}[h]
\centering
\includegraphics[width=8.9cm]{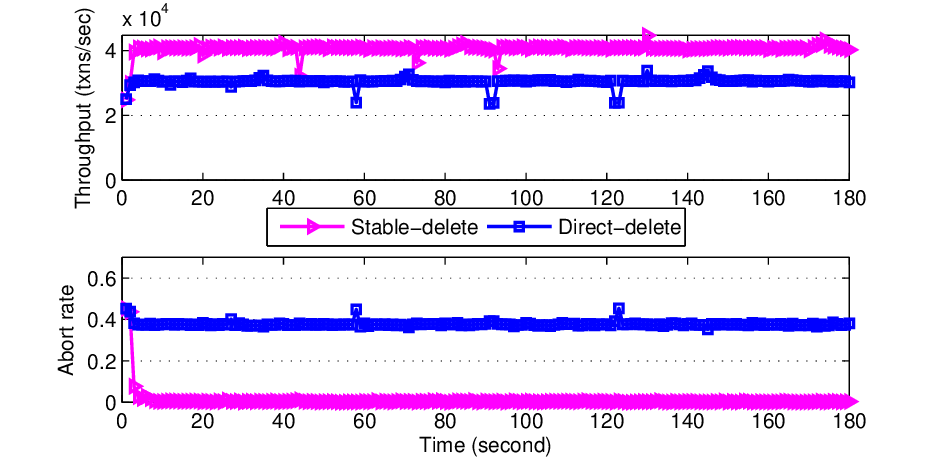}
\caption{The comparison of Stable-delete and Direct-delete. This graph shows why we design stable-delete in our system. If the items are deleted directly, there is no enough information to decide which queue (thread) is supposed to use. Then transactions are inserted into the same queue, that leads to high abort rate.}
\label{fig:delete} %% label for entire figure
\end{figure}

As we can see, there is a memory spike when the DBMS start to run. That is because only when the system enters stable mode, it starts to delete entries in the state table. So in the graph, before 10000, the system still have high abort rate, means system is not stable. In this experiment, we turn off this feature and the performance becomes bad~\cref{fig:delete}

To run continuously, the system must adapt to two circumstances. The first is memory pressure. The History and State components continually allocate new memory and without some management these components will consume all the memory in the system. The History is simply periodically regenerated from scratch, thus reclaiming its memory and refreshing the distribution recorded. But the State needs more fine grained control of the entries in the table. The access rate (R), total references (T) and the queue with the maximum references (Q) are used as criteria to selectively remove entries - essentially references are remove if they have low access rate, few total references, and belong to a queue with many references.

\begin{figure*}[h]
\centering
\includegraphics[width=18.9cm]{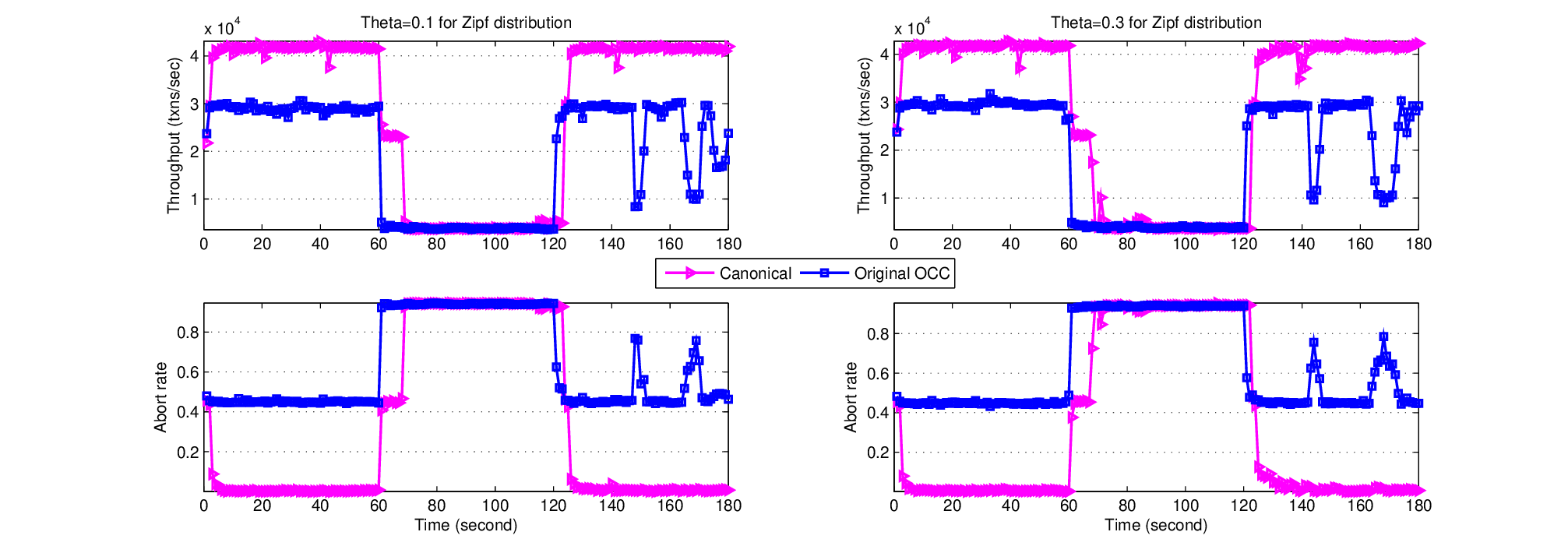}
\caption{\textbf{Performance of OCC over Time} -- The performance of the system as the workload distribution changes. The graph shows the performance of Original OCC and Canonical. For the first and last 60 seconds the workload distribution is uniform for 20 warehouses. For the middle 60 seconds the TPC-C workload distribution is skewed with Theta=0.1 or 0.3}
\label{fig:adaptive13} %% label for entire figure
\end{figure*}

The second circumstance is the shift in distribution of the workload. We  show the result of adaptive running in~\cref{fig:adaptive13} and ~\cref{fig:adaptive69}. "Adaptive" means the distribution of transactions changes during running. In this experiment, the warehouse distribution is uniform distribution during the first 60 seconds. Then the distribution  changes to Zipf distribution among the warehouse for the second 60 seconds. After that, it switches back to uniform distribution. Figure~\cref{fig:adaptive13} and ~\cref{fig:adaptive69} shows the results for different Zipf theta values, where a higher theta value introduces more skew. 

The figures illustrate that during the first 60 seconds, since it is uniform distribution, the throughput and abort rate are stable for both Canonical and Original OCC. 
When the distribution changes to distribution after the first 60 seconds, the throughput of Original OCC drops very quickly, because for Zipf distribution, more transactions belongs to the same warehouse, which lead to more conflict. For Canonical, the performance drops as well, but not as quickly as Original OCC. That is because the transactions belonging to the same warehouse are still assigned to the same queue. But it can't continue to a long time, since when a queue is empty, stealing occurs. 
When the distribution switches back to uniform  (after 120 seconds), the adaption mechanism works. 
%Canonical will try to do the algorithm after a while when the abort rate is above a threshold. But for Original OCC, it only randomly assigns transactions. Therefore, after the distribution switches back, the imbalance in the queue lengths remains in randomized scheduling for considerable time. This imbalance causes some threads to be idle and then steal transactions, increasing the abort rate.
%\todo{first this shows that we do poorly with a zipfian distribution. second, why is it not table?}

%%%%%%%% adaptive %%%%%%%%%%

%\begin{figure*}[h]
%\centering
%\includegraphics[width=18.9cm]{fig/adaptive_06_09.eps}
%\caption{Adatpive running with Theta=0.6 and 0.9}
%\label{fig:adaptive69} %% label for entire figure
%\end{figure*}

\subsection{Instability}
Our experimental results investigated a broad range of benchmarks and every large number of design choices. 
One question that arose in general during our analysis is the sensitivity of our policies to the initial conditions of the State table. 
This sensitivity occurs because the State component records the first few transaction in queues and subsequent queuing decisions are based on those results. 
%In part this fact leads to the consequences of the performance impact of the Balance policy described in the previous section. 
This instability is also in part a consequence of the benchmarks having a small range of queries.
In future work, we will consider methods to reduce this instability.

\subsection{Robustness}
Robustness is a system exhibiting low variance of the performance of 
a policy when subjected to minor variations in the workload. In our experiments, we ran each experiment ten times and reported the average measurement. However, our relative standard error is below 1\% for our recommended Canonical policy, generally matching the relative error of Hard Partition, indicating that the policy is relatively robust.
This robustness is due in large part to the State component maintaining an aggregated history of the references appearing in different queues. 
After a few thousand processed transactions this state becomes relatively stable because popular references will have data associated with them.

%% ==================================================================
%% Related Work
%% ==================================================================
\section{Related Work}
\label{sec:related}

In this section, we discuss various work related to our approach.
\\ \vspace{-0.05in}

\textbf{Semantic analysis}.
A deterministic approach to the analysis of code is followed in~\cite{Manjhi:2009}. This work applies two heuristics to deterministic analysis of application code to improve query processing. One approach, termed holistic query processing, merges database calls in an application code loop back into a single SQL query. The second identifies dependencies between database calls and extracts the independent statements to be run in parallel to decrease latency. A similar approach, extended to use optimization, is followed in QURO~\cite{Yan:2016:VLDB}, to reorder statements to reduce transaction conflict under a two-phase locking protocol. It formulates the reordering problem using Integer Linear Programming to put the most contentious queries at the end of the transaction. 
% don't insult other others
%However, reordering requires complex analysis for query dependency which brings expensive cost for query parsing. Further more, it can't reorder the queries when they have dependencies for each other and therefore can't handle dependency transactions.  
These papers and our approach both use transactional analysis to improve performance. The main difference is that our approach analyzes transactions arriving at the server and not in client code.

%[bailis, calvin] compare: both are detecting conflict, either deterministically or (us) statistically. both then schedule transactions deterministically or (us) 'statistically'. contrast: bailis/calvin have high engineering costs - for the language uses, either deep knowledge of logical proofs or the read/write sets must be known. In our case, little engineering work is required - some knowledge of the possible features are needed, but the history of conflicts itself is used to statistically model conflicts, instead of deep analysis.

%the computational and engineering costs of deterministic analysis are high. the computational and engineering costs of statistical analysis are low. however, the fundamental assumption of statistical analysis is something like an 80/20 rule: we can cheaply determine conflict 80 percent of the time and we give up on highly detailed modeling to capture the other twenty percent.

In~\cite{Harchol-Balter:2003}, the performance impact of scheduling based on the size of static HTTP requests to a server is investigated. 
The paper improves server response time by scheduling draining network socket buffers according to Shortest Remaining Processing Time queuing policy instead of Round Robin or random. 
Priorities on the queues are set by the number of bytes left to read from a static file.
The paper and our work both use scheduling to improve performance based on a feature that characterizes the work remaining to be done for a an operation.
The main difference is that the feature is a property of the transaction in the paper, where as in our work the feature is a signal generated by the transaction processing system.

Ic3~\cite{Wang:2016:SMD:2882903.2882934} chops transactions into pieces and analyzes the dependency between transaction pieces. Ic3 uses chopping technique, that is introduced by~\cite{Shasha:1995:TCA:211414.211427}, to construct a dependency graph and maintains the dependency for running transactions. 
% Ic3's purpose is to provide efficient serializability, while our method can be deployed on top of any concurrency control protocol. Further more, to control the overhead, ic3 only tracks the direct dependency, and a transaction without such analysis must wait for all dependent transactions to complete, that could increase response time for such kind of transactions. 
The main similarity between this work and our approach is that both analyze dependencies transactions to improve performance. However, our dependency structure is implied in the statistical relationship between references that occur in transactions, as opposed to an explicit dependency model.
\\ \vspace{-0.05in}

\textbf{Scheduling}.
Some researchers utilize hardware to accelerate transaction processing~\cite{hardware1,hardware2,hardware3,hardware4,hardware5}. But they do not address contention issue using scheduling methods. In\cite{Blake:2009}, scheduling for hardware transactional memory is studied. Transaction boundaries in this context are annotated in source code. The program counter is used as the transaction identifier. The paper describes a scheduling system that maintains a global data structure that records a history of conflicts between transactions and information about each transaction.
When a processor is ready to schedule a thread, it examines the structure and decides to execute, stall briefly, or swap in a new thread. 
Both this work and our approach use the history of transaction conflicts. 
The paper describes a system that records much more complex additional information about each transaction. When scheduling a new transaction, the system computes the probability of conflict between the new transactions and each existing currently executing transaction. In our work, we track, essentially, only the input parameters to transactions and then schedule transactions into queues.
In addition the paper computes a metric of a pair-wise probability of abort between two transaction. Although direct contrasts between the two works are difficult, the results the previous section of this paper suggest that simple global counting methods should be studied as an alternative to probabilistic models.
\\ \vspace{-0.05in}
%Another complementary approach to improving database performance focuses on application program analysis instead of our focus on backend database improvement. One area of application program analysis focuses on improvement performance (network latency in particular) through heuristic program analysis~\cite{Manjhi:2009}

%The basic idea of Sloth is quite clever - they apply lazy evaluation semantics to the database application code and database calls that appear in an application. Essentially the database calls are batched together through lazy evaluation and then shipped to the dbms. At that point, even trivially one could simply evaluate all the batched queries and ship all the results back in one message, thus reducing network latency round trips. (This work is a generalization to some earlier work Amit Manjhi did with a group of us.) 

%The work is complementary to ours since it is focused on transformation of the application to improve performance as opposed to our backend approach.

\textbf{Data and functionality partitioning}. In H-Store~\cite{Stonebraker:2007:EAE:1325851.1325981}, data is partitioned through analyzing store procedure and each thread corresponds a partition. The transactions execute serially for a given partition. HyPer~\cite{hyper:icde11} also processes transactions using a single-thread. It supports both OLTP and OLAP workloads (the latest version of Hyper uses optimistic multi-versioning~\cite{hyper:sigmod15}). In DORA~\cite{Pandis:2010:DTE:1920841.1920959}, a transaction is splitted up into sub-transactions and a sub-transaction is assigned to a single partition to execute. 

Cheung et al.~propose a technique of automatic code partitioning using program analysis~\cite{Cheung:2012:APD:2350229.2350262}. By code partitioning they mean the partitioning of the application program between the application and stored procedures on the DBMS. The paper basically does a program analysis of the application and then chops up the program into pieces with respect to network latency. Some pieces are executed on the application server and other pieces are executed on the database server. In effect, store procedures are automatically defined through program analysis. Our method does not analyze code. Instead we separate a functionality to analyze conflicts and assign transactions with different conflicts to the concurrency control protocol to execute. 
\\ \vspace{-0.05in}

\textbf{Multi-core optimization}. Larson et al.~propose an optimistic validation protocol that avoids using global critical section~\cite{Larson:2011:HCC:2095686.2095689}. Their methods address the scalability problem for both pessimistic and optimistic concurrency control. In Silo~\cite{Tu:2013:STM:2517349.2522713}, an optimistic in-memory multi-core DBMS is proposed to avoids centralized contention points. It provides a commit protocol with serializability and eliminates all shared-memory writes. Johnson et al. use speculative lock inheritance to decrease the frequency of contended latch acquisitions~\cite{Johnson:2009:IOS:1687627.1687682}. Their method passes contended logical locks without invoking the lock manager. All of these systems focus the concurrency control protocol design. In contrast, our approach builds another layer that is on top of concurrency control and can be used by any concurrency control protocol. 

ORTHRUS~\cite{Ren:2016:DPS:2882903.2882958} separates execution threads from concurrency control threads. Execution threads send message to concurrency control threads to acquire locks for accessed records. If a transaction accesses several records, the execution thread might send message to different concurrency control threads. In order to fully utilize CPU cycles, concurrency control threads may send message to other concurrency control thread to help acquiring locks. The difference is that our paper focuses on grouping conflicts and decoupling from concurrency control protocol while ORTHRUS breaks concurrency control into two functionality modules.

%Pavlo CIDR 2017 paper 

%recent work in workload modeling promising

%document work on tools to do: automatic index selection, other issues.

\section{Conclusion}
\label{sec:conlusion}

%Database administrators (DBAs) have a complex and demanding job. They must understand the application requirements of a large collection of application program and how these programs function over time and resources. In addition, DBAs must understand the internal functioning of a database system. Then, DBAs must integrate this knowledge to effectively tune databases to achieve higher performance. Self-driving database research aims to replace part of the responsibilities of DBAs with autonomous decision making systems. 

In this paper, we provide preliminary evidence that scheduling of  OLTP transactions offers both flexibility (non-partitioned) and performance. The key is that transaction scheduling can be improved by observing the history of transaction aborts and integrating this information into a design that improves transaction throughput, under high load, without significantly impacting response time. 
The design is remarkably similar to data partitioning, a technique that is the basis of an entire class of OLTP architectures.
The method and its supporting system is designed to have a low engineering footprint and it is amenable to implementation either external or internal to a main-memory database system. The design works equally well for optimistic concurrency control protocols as well as pessimistic concurrency control protocols based on strict 2-phase locking.

In future work, a direct extension of this work would handle a larger class of predicates, including inequalities, range queries, and string matching operations. More broadly, we will explore a broader set of workloads.  In particular, online analytic processing (OLAP) queries present a challenge because the read/write sets of a query are more complex than the read/write sets of a OLTP transaction. In addition, hybrid transaction and analytic processing (HTAP) workloads present the additional challenge of combining the complexity of query processing with the conflicts of transactions. 
%The current favored solution for HTAP is to abandon serializability and offer multiversioned results. Perhaps this solution can be replaced with intelligent scheduling. 

% \section{Acknowledgements}
% ...

\bibliographystyle{abbrv}
\bibliography{ref}

\balancecolumns 

%\appendix

\section{Appendix}
\begin{table}[h]
  \centering
  \begin{tabular}{lr}
  \multicolumn{2}{c}{TPC-C} \\
  parameter & count \\ \hline
warehouse & 11 \\
district & 110 \\
item & 10,000\\
customer  & 33,000\\
history & 33,000\\
stock & 110,000\\
orders & 330,00\\
new order & 33,000\\
order line & 330,030\\
    \end{tabular}
\begin{tabular}{lr}
  \multicolumn{2}{c}{Smallbank} \\
  parameter & count \\ \hline
account & 10,000 \\
savings & 10,000 \\
checking & 10,000\\
\end{tabular}
    \begin{tabular}{lr}
  \multicolumn{2}{c}{TATP} \\
  parameter & count \\ \hline
subscriber & 10,000 \\
access info & 24,986 \\
special facility & 24,931\\
call forwarding  & 37,333\\
    \end{tabular}
  \caption{Experiment parameters that determine the initial configuration of the database at the start of an experiment. }
  \label{tab:parameters}
\end{table}

\begin{table}[h]
  \centering
  \begin{tabular}{lr}
  \multicolumn{2}{c}{TPC-C} \\
  transaction & \% \\ \hline
NewOrder & 50 \\
Payment & 50 \\
    \end{tabular}
\begin{tabular}{lr}
  \multicolumn{2}{c}{Smallbank} \\
  transaction & \% \\ \hline
Amalgamate & 4 \\
Balance & 24 \\
DepositChecking & 24\\
TransactSavings & 24\\
WriteCheck & 24 \\ \hline
account hot spot 90\% & [1-50]\\
\end{tabular}
    \begin{tabular}{lr}
  \multicolumn{2}{c}{TATP} \\
  transaction & \% \\ \hline
 GetAccessData & 3\\
 GetNewDestination & 3 \\
 GetSubscriberData & 40 \\
 InsertCallForwarding & 2 \\  
 DeleteCallForwarding & 2 \\
 UpdateSubscriberData & 10 \\
 UpdateLocation & 40 \\ \hline
 subscriber hot spot 90\% & [1-20]\\
    \end{tabular}
  \caption{Experiment workload. The hot spot range was added to increase the conflict rate.}
  \label{tab:workload}
\end{table}

\end{document}